\documentclass{article}
\usepackage[english]{babel}
\usepackage[round]{natbib}
\usepackage[printonlyused]{acronym}	
\usepackage[section]{placeins} 
\usepackage[tracking=true]{microtype}  
\usepackage{amsmath} 
\usepackage{mathtools} 
\usepackage{booktabs} 
\usepackage{siunitx} 
\usepackage{multirow} 

\usepackage{subfig}
\usepackage{tikz}
\usetikzlibrary{arrows}	
\usetikzlibrary{shapes}	
\usepackage{threeparttable}
\usepackage{chngpage} 

\usepackage[linesnumbered,ruled,vlined]{algorithm2e}

\SetCommentSty{mycommfont} 

\usepackage{afterpage} 

\newcolumntype{H}{>{\setbox0=\hbox\bgroup}c<{\egroup}@{}} 

\usepackage[tracking=true]{microtype}

\usepackage[hidelinks]{hyperref}

\begin{document}
\title{A large neighbourhood based heuristic for two-echelon routing problems}
\author{\textbf{Ulrich~Breunig}\\Department of Business Administration\\University of Vienna, Austria\\ulrich.breunig@univie.ac.at\\ \\ \textbf{Verena~Schmid}\\Christian Doppler Laboratory for Effcient Intermodal Transport Operations\\University of
Vienna, Austria\\verena.schmid@univie.ac.at\\ \\\textbf{Richard~F.~Hartl}\\Department of Business Administration\\University of Vienna, Austria\\richard.hartl@univie.ac.at\\ \\\textbf{Thibaut~Vidal}\\Departemento de Informática\\Pontifícia Universidade Católica do Rio de Janeiro, Brazil\\thibaut.vidal@mit.edu}





\maketitle

\begin{abstract}
In this paper, we address two optimisation problems arising in the context of city logistics and two-level transportation systems. The \acl{2E-VRP} and the \acl{2E-LRP} seek to produce vehicle itineraries to deliver goods to customers, with transits through intermediate facilities.
To efficiently solve these problems, we propose a hybrid metaheuristic which combines enumerative local searches with destroy-and-repair principles, as well as some tailored operators to optimise the selections of intermediate facilities. We conduct extensive computational experiments to investigate the contribution of these operators to the search performance, and measure the performance of the method on both problem classes. The proposed algorithm finds the current best known solutions, or better ones, for 95\% of the \acl{2E-VRP} benchmark instances. Overall, for both problems, it achieves high-quality solutions within short computing times. Finally, for future reference, we resolve inconsistencies between different versions of benchmark instances, document their differences, and provide them all online in a unified format.
\end{abstract}

\section{Introduction}

The traffic of vehicles is a major nuisance in densely populated areas. Trucks disturb peoples' well-being by emitting noise and air pollution. As the amount of goods in transit increases, a proper planning of road networks and facility locations becomes critical to mitigate congestion.
To face these challenges, algorithmic tools have been developed to optimise city logistics at several levels: considering traffic regulation, itineraries and network design choices. Boosting the efficiency of goods transportation from suppliers to customers presents important challenges for different planning horizons. On the operational level, efficient itineraries must be found for the available vehicles from day to day, \acs{eg}, reducing the travelled distance. On a tactical level, the overall fleet size, vehicle dimensions, capacities and characteristics are of interest. Larger trucks are more efficient in terms of cost per shipped quantity, whereas smaller vehicles are more desirable in city centres: they emit less noise, and only need smaller parking spots. Finally, the clever selection of locations for production sites, warehouses, and freight terminals is a typical strategic decision. 

In this article, we consider the problem of jointly determining good routes to deliver goods to customers, and at which intermediate facilities a switch from larger trucks to smaller city freighters should happen. This problem is challenging, due to the combination of these two families of combinatorial decisions. 

To address this problem, we propose a simple metaheuristic, which combines local and large neighbourhood search with the ruin and recreate principle.
The method is conceptually simple and fast, exploiting a limited subset of neighbourhoods in combination with a simple strategy for closing and opening intermediate facilities. 
We conduct extensive computational experiments on \ac{2E-VRP} and \ac{2E-LRPSD} instances to investigate the contribution of these operators, and measure the performance of the method on both problem classes. For the \ac{2E-VRP}, this algorithm is able to reach or outperform $95\%$ of the best known solutions. In general, for both problems, high-quality solutions are attained in short computing times. As such, this algorithm will serve as a good basis for future developments on more complex and realistic two-tiered delivery problems.

The paper is organised as follows. The problems are described in Section~\ref{description} and an overview of the related literature is given in Section~\ref{literature}. Mathematical formulations are presented in Section~\ref{mathmodel}. Section~\ref{method} describes the proposed algorithm. Section~\ref{experiments} reviews current available benchmark instances and examines the performance of the proposed method. Section~\ref{conclusion} concludes.
\section{Problem description}
\label{description}
\acp{Vrp} \acused{vrp} are a class of combinatorial optimisation problems, which aim to find good itineraries to service a number of customers with a fleet of vehicles.
The \ac{2E-VRP} is a variant of a \ac{vrp}, which exploits the different advantages of small and large vehicles in an integrated delivery system. The goal is to design an efficient distribution chain, organised in two levels: trucks operate on the first level between a central depot and several selected intermediate distribution facilities, called satellites. The second level also includes the satellites -- because both levels are interconnected there -- as well as the end-customers. Small city freighters are operated between satellites and customers. The depot supplies sufficient quantities to satisfy all customer demands. The products are directly transferred from trucks to city freighters at satellite locations. These city freighters will perform the deliveries to the final customers. Any shipment or part of shipment has to transit through exactly one satellite, and the final delivery to the customer is done in one block. As such, split deliveries are not allowed for city freighters. The quantity (``demand'') of goods shipped to a satellite is not explicitly given, but evaluated as the sum of all customer demands served with city freighters originating from this satellite. Depending on the second level itineraries, split deliveries can occur on the first level since the total quantity needed at one satellite can exceed the capacity of one truck.

Finding good combined decisions for routing and intermediate facility openings is significantly more difficult than in well-studied settings such as the \ac{CVRP}. The special case of a \ac{2E-VRP} with only one satellite can be seen as a \ac{vrp} \citep{Cuda2014a, Perboli2011}. The first level of the \ac{2E-VRP} reduces to a \ac{CVRP} with split deliveries. The structure of the second level is a \ac{MD-VRP}, where the depots correspond to the satellite locations \citep{Jepsen2012}. Those two levels have to be synchronised with each other. The \ac{2E-VRP} is a generalisation of the classical \ac{vrp} and is thus NP-hard.

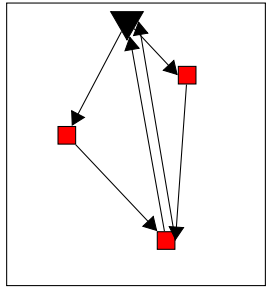
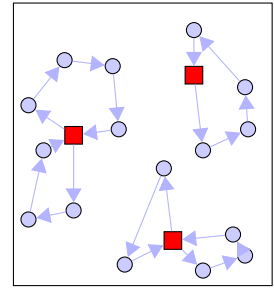
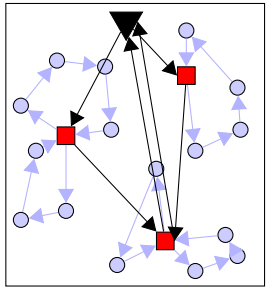
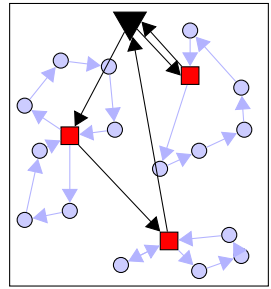
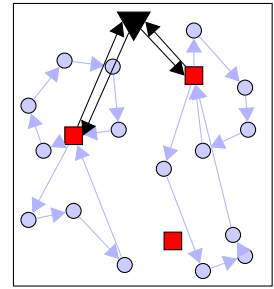
\begin{figure}[htb]
	\subfloat[\ac{SD-VRP} / first level of \ac{2E-VRP}]{
	\label{graph_sdvrp}
		\begin{tikzpicture}[scale=0.04, every node/.style={font=\tiny},>=triangle 60, depot/.style={rectangle,draw,fill=red},
customer/.style={circle,draw,fill=blue!20,inner sep=0pt,minimum size=2mm}]
		\node (Depot)  at (-5,43)  [regular polygon,regular polygon sides=3,draw,fill=black,inner sep=0pt,minimum size=5mm,rotate=180] {};
		
		\draw (-45,-45) rectangle (41,49);

		\node (D1)  at (-25,5)  [depot] {};
		\node (D2)  at ( 15,25) [depot] {};
		\node (D3)  at ( 8,-30) [depot] {};

		\draw [->] (Depot) -- (D1);
		\draw [->] (D1) -- (D3);
		\draw [->] (D3) -- (Depot);
		
		\draw [->] (Depot) -- (D2.west);
		\draw [->] (D2) -- (D3.east);
		\draw [->] (D3.east) -- (Depot.west);
	\end{tikzpicture}
	} 
	\hfill
	\subfloat[\ac{MD-VRP} / second level of \ac{2E-VRP}]{
	\label{graph_mdvrp}
		\begin{tikzpicture}[scale=0.04, every node/.style={font=\tiny},>=triangle 60, depot/.style={rectangle,draw,fill=red},
	customer/.style={circle,draw,fill=blue!20,inner sep=0pt,minimum size=2mm}]
	
	\draw (-45,-45) rectangle (41,49);
	

			\node (D1)  at (-25,5)  [depot] {};
			\node (A)   at (-40,15) [customer] {};
			\node (B)   at (-28,30) [customer] {};
			\node (C)   at (-12,28) [customer] {};
			\node (D)   at (-10,7)  [customer] {};
			\node (E)   at (-25,-20)[customer] {};
			\node (F)   at (-40,-23)[customer] {};
			\node (G)   at (-35,0)  [customer] {};
			\node (D2)  at ( 15,25) [depot] {};
			\node (H)   at ( 18,0)  [customer] {};
			\node (I)   at ( 33,7)  [customer] {};
			\node (J)   at ( 32,21) [customer] {};
			\node (K)   at ( 15,40) [customer] {};
			\node (D3)  at ( 8,-30) [depot] {};
			\node (L)   at ( 5,-6)  [customer] {};
			\node (M)   at (-8,-38) [customer] {};
			\node (N)   at (18,-40) [customer] {};
			\node (O)   at (32,-35) [customer] {};
			\node (P)   at (28,-28) [customer] {};
			\draw [blue!30,->] (D1) -- (A);
			\draw [blue!30,->] (A) -- (B);
			\draw [blue!30,->] (B) -- (C);
			\draw [blue!30,->] (C) -- (D);
			\draw [blue!30,->] (D) -- (D1);
			\draw [blue!30,->] (D1) -- (E);
			\draw [blue!30,->] (E) -- (F);
			\draw [blue!30,->] (F) -- (G);
			\draw [blue!30,->] (G) -- (D1);
			\draw [blue!30,->] (D2) -- (H);
			\draw [blue!30,->] (H) -- (I);
			\draw [blue!30,->] (I) -- (J);
			\draw [blue!30,->] (J) -- (K);
			\draw [blue!30,->] (K) -- (D2);
			\draw [blue!30,->] (D3) -- (L);
			\draw [blue!30,->] (L) -- (M);
			\draw [blue!30,->] (M) -- (D3);
			\draw [blue!30,->] (D3) -- (N);
			\draw [blue!30,->] (N) -- (O);
			\draw [blue!30,->] (O) -- (P);
			\draw [blue!30,->] (P) -- (D3);	
		\end{tikzpicture}
	} 
	 
	\subfloat[\ac{2E-VRP} I]{
	\label{graph_2esplit}
	\begin{tikzpicture}[scale=0.04, every node/.style={font=\tiny},>=triangle 60, depot/.style={rectangle,draw,fill=red},
customer/.style={circle,draw,fill=blue!20,inner sep=0pt,minimum size=2mm}]
		\node (Depot)  at (-5,43)  [regular polygon,regular polygon sides=3,draw,fill=black,inner sep=0pt,minimum size=5mm,rotate=180] {};
		
		\draw (-45,-45) rectangle (41,49);

		\node (D1)  at (-25,5)  [depot] {};
		
		\node (A)   at (-40,15) [customer] {};
		\node (B)   at (-28,30) [customer] {};
		\node (C)   at (-12,28) [customer] {};
		\node (D)   at (-10,7)  [customer] {};
		
		\node (E)   at (-25,-20)[customer] {};
		\node (F)   at (-40,-23)[customer] {};
		\node (G)   at (-35,0)  [customer] {};
		
		\node (D2)  at ( 15,25) [depot] {};
		
		\node (H)   at ( 18,0)  [customer] {};
		\node (I)   at ( 33,7)  [customer] {};
		\node (J)   at ( 32,21) [customer] {};
		\node (K)   at ( 15,40) [customer] {};
		
		\node (D3)  at ( 8,-30) [depot] {};
		
		\node (L)   at ( 5,-6)  [customer] {};
		\node (M)   at (-8,-38) [customer] {};
		
		\node (N)   at (18,-40) [customer] {};
		\node (O)   at (32,-35) [customer] {};
		\node (P)   at (28,-28) [customer] {};

		\draw [blue!30,->] (D1) -- (A);
		\draw [blue!30,->] (A) -- (B);
		\draw [blue!30,->] (B) -- (C);
		\draw [blue!30,->] (C) -- (D);
		\draw [blue!30,->] (D) -- (D1);
		\draw [blue!30,->] (D1) -- (E);
		\draw [blue!30,->] (E) -- (F);
		\draw [blue!30,->] (F) -- (G);
		\draw [blue!30,->] (G) -- (D1);
		\draw [blue!30,->] (D2) -- (H);
		\draw [blue!30,->] (H) -- (I);
		\draw [blue!30,->] (I) -- (J);
		\draw [blue!30,->] (J) -- (K);
		\draw [blue!30,->] (K) -- (D2);
		\draw [blue!30,->] (D3) -- (L);
		\draw [blue!30,->] (L) -- (M);
		\draw [blue!30,->] (M) -- (D3);
		\draw [blue!30,->] (D3) -- (N);
		\draw [blue!30,->] (N) -- (O);
		\draw [blue!30,->] (O) -- (P);
		\draw [blue!30,->] (P) -- (D3);
		\draw [->] (Depot) -- (D1);
		\draw [->] (D1) -- (D3);
		\draw [->] (D3) -- (Depot);
		
		\draw [->] (Depot) -- (D2.west);
		\draw [->] (D2) -- (D3.east);
		\draw [->] (D3.east) -- (Depot.west);
	\end{tikzpicture}	
	}
	\hfill
	\subfloat[\ac{2E-VRP} II]{
	\label{graph_2e}
	\begin{tikzpicture}[scale=0.04, every node/.style={font=\tiny},>=triangle 60, depot/.style={rectangle,draw,fill=red},
customer/.style={circle,draw,fill=blue!20,inner sep=0pt,minimum size=2mm}]
		\node (Depot)  at (-5,43)  [regular polygon,regular polygon sides=3,draw,fill=black,inner sep=0pt,minimum size=5mm,rotate=180] {};
		
		\draw (-45,-45) rectangle (41,49);

		\node (D1)  at (-25,5)  [depot] {};
		
		\node (A)   at (-40,15) [customer] {};
		\node (B)   at (-28,30) [customer] {};
		\node (C)   at (-12,28) [customer] {};
		\node (D)   at (-10,7)  [customer] {};
		
		\node (E)   at (-25,-20)[customer] {};
		\node (F)   at (-40,-23)[customer] {};
		\node (G)   at (-35,0)  [customer] {};
		
		\node (D2)  at ( 15,25) [depot] {};
		
		\node (H)   at ( 18,0)  [customer] {};
		\node (I)   at ( 33,7)  [customer] {};
		\node (J)   at ( 32,21) [customer] {};
		\node (K)   at ( 15,40) [customer] {};
		
		\node (D3)  at ( 8,-30) [depot] {};
		
		\node (L)   at ( 5,-6)  [customer] {};
		\node (M)   at (-8,-38) [customer] {};
		
		\node (N)   at (18,-40) [customer] {};
		\node (O)   at (32,-35) [customer] {};
		\node (P)   at (28,-28) [customer] {};

		\draw [blue!30,->] (D1) -- (A);
		\draw [blue!30,->] (A) -- (B);
		\draw [blue!30,->] (B) -- (C);
		\draw [blue!30,->] (C) -- (D);
		\draw [blue!30,->] (D) -- (D1);
		\draw [blue!30,->] (D1) -- (E);
		\draw [blue!30,->] (E) -- (F);
		\draw [blue!30,->] (F) -- (G);
		\draw [blue!30,->] (G) -- (D1);
		\draw [blue!30,->] (D2) -- (L);
		\draw [blue!30,->] (L) -- (H);
		\draw [blue!30,->] (H) -- (I);
		\draw [blue!30,->] (I) -- (J);
		\draw [blue!30,->] (J) -- (K);
		\draw [blue!30,->] (K) -- (D2);

    \draw [blue!30,->] (D3) -- (M);
		\draw [blue!30,->] (M) -- (D3);
		\draw [blue!30,->] (D3) -- (N);
		\draw [blue!30,->] (N) -- (O);
		\draw [blue!30,->] (O) -- (P);
		\draw [blue!30,->] (P) -- (D3);
		\draw [->] (Depot) -- (D1);
		\draw [->] (D1) -- (D3);
		\draw [->] (D3) -- (Depot);
		\draw [->] (Depot) -- (D2.west);
		\draw [->] (D2) -- (Depot.west);
	\end{tikzpicture}
	}
	\hfill
	\subfloat[\ac{2E-VRP} III]{
	\label{graph_2e_notall}
		\begin{tikzpicture}[scale=0.04, every node/.style={font=\tiny},>=triangle 60, depot/.style={rectangle,draw,fill=red},
customer/.style={circle,draw,fill=blue!20,inner sep=0pt,minimum size=2mm}]
		\node (Depot)  at (-5,43)  [regular polygon,regular polygon sides=3,draw,fill=black,inner sep=0pt,minimum size=5mm,rotate=180] {};
		
		\draw (-45,-45) rectangle (41,49);

		\node (D1)  at (-25,5)  [depot] {};
		
		\node (A)   at (-40,15) [customer] {};
		\node (B)   at (-28,30) [customer] {};
		\node (C)   at (-12,28) [customer] {};
		\node (D)   at (-10,7)  [customer] {};
		
		\node (E)   at (-25,-20)[customer] {};
		\node (F)   at (-40,-23)[customer] {};
		\node (G)   at (-35,0)  [customer] {};
		
		\node (D2)  at ( 15,25) [depot] {};
		
		\node (H)   at ( 18,0)  [customer] {};
		\node (I)   at ( 33,7)  [customer] {};
		\node (J)   at ( 32,21) [customer] {};
		\node (K)   at ( 15,40) [customer] {};
		
		\node (D3)  at ( 8,-30) [depot] {};
		
		\node (L)   at ( 5,-6)  [customer] {};
		\node (M)   at (-8,-38) [customer] {};
		
		\node (N)   at (18,-40) [customer] {};
		\node (O)   at (32,-35) [customer] {};
		\node (P)   at (28,-28) [customer] {};
		
		\draw [blue!30,->] (D2) -- (K);
		\draw [blue!30,->] (K) -- (J);
		\draw [blue!30,->] (J) -- (I);
		\draw [blue!30,->] (I) -- (H);
		\draw [blue!30,->] (H) -- (D2);
		
		\draw [blue!30,->] (D2) -- (L);
		\draw [blue!30,->] (L) -- (N);
		\draw [blue!30,->] (N) -- (O);
		\draw [blue!30,->] (O) -- (P);
		\draw [blue!30,->] (P) -- (D2);
		
		\draw [blue!30,->] (D1) -- (F);
		\draw [blue!30,->] (F) -- (E);
		\draw [blue!30,->] (E) -- (M);
		\draw [blue!30,->] (M) -- (D1);
		
		\draw [blue!30,->] (D1) -- (G);
		\draw [blue!30,->] (G) -- (A);
		\draw [blue!30,->] (A) -- (B);
		\draw [blue!30,->] (B) -- (C);		
		\draw [blue!30,->] (C) -- (D);
		\draw [blue!30,->] (D) -- (D1);	
		\draw [->] (Depot) -- (D1.east);
		\draw [->] (D1) -- (Depot.east);
		
		\draw [->] (Depot) -- (D2.west);
		\draw [->] (D2) -- (Depot.west);
	\end{tikzpicture}
	}
	\caption{Subproblems related to the \acs{2E-VRP} and different solutions depending on which intermediate facilities are used}
	\label{graph_solutions}
\end{figure}

Figure \ref{graph_solutions} shows different set-ups for goods distribution. The depot is represented by a triangle, satellites by squares, and customers by circles. Figures~\ref{graph_sdvrp} and \ref{graph_mdvrp} show graphical representations of a \ac{SD-VRP} and a \ac{MD-VRP} respectively. Figures \ref{graph_2esplit}--\ref{graph_2e_notall} represent feasible solutions for the \ac{2E-VRP}: with split deliveries occurring at one of the satellites, without split deliveries, and in Figure \ref{graph_2e_notall} a solution where only a subset of satellites is used.

The proposed algorithm has primarily been designed for the \ac{2E-VRP}, and then tested on the \ac{2E-LRPSD}, which includes additional tactical decisions. 
The basic structure of the \ac{2E-LRPSD} is very similar to the \ac{2E-VRP}. The main difference is that it corresponds to a more tactical planning since only potential locations for depots or satellites are known and the use of any location results in opening costs. In contrast with the \ac{2E-VRP}, the fleet size is unbounded, but fixed costs are counted for the use of each vehicle. The classical benchmark sets from the literature include different costs per mile for large first level trucks and small city freighters, unlike in \ac{2E-VRP} benchmark instances, where mileage costs are identical for all vehicles. Finally, split deliveries are not allowed at both levels.
We thus applied our algorithm to \ac{2E-LRPSD} instances, where location decisions have to be taken at the secondary facilities. Following the notations of \cite{boccia2011location} we focus on $3/T/\overline{T}$ problems.
\section{Literature Review}
\label{literature}
\cite{Jacobsen1980} were amongst the first to introduce a two-echelon distribution optimisation problem. They proposed a three stage heuristic to solve the daily distribution of newspapers in Denmark, but no mathematical model was designed. Several possible transfer points were considered to transfer newspapers from one vehicle to another. The solution to this problem consists of three layers of decisions: the number and location of transfer points, the tours from the printing office and the tours from the transfer points to the retailers. An improved solution algorithm for the same problem can be found in \cite{Madsen1983}. Following the nomenclature of the authors and the classification in the recent survey on two-echelon routing problems by \cite{Cuda2014a}, the problem includes location decisions. Nevertheless, from today's point of view, it cannot be categorised as a \ac{2E-LRP} as there are no opening costs associated with the use of intermediate facilities and retailers can also be served directly from the printing office without using intermediate nodes. In addition, the problem cannot be categorised as a \ac{2E-VRP} since first-level split deliveries are not allowed.

\paragraph{Two-echelon vehicle routing problem}
\cite{crainic2004advanced} used data from Rome to study an integrated urban freight management system. As large trucks cannot pass through the narrow streets in the city centre, they used intermediate facilities to redistribute loads from large trucks to smaller vehicles. The city was divided into several commercial and external zones, and a mathematical location-allocation formulation was proposed and solved using a commercial solver. A comparison between solutions for delivering goods lead to the conclusion that intermediate facilities reduce the use of large trucks significantly, and more work is done by smaller city freighters. 

\cite{crainic2009models} formulated a time dependent version of the problem, including time windows at the customers. To our knowledge, there are no test instances or solution approaches for this variant so far. \cite{crainic2010two} studied the impact of different two-tiered transportation set-ups on total cost. According to their results, the \ac{2E-VRP} can yield better solutions than the \ac{vrp} if the depot is not located within the customer area but externally. \cite{Perboli2011} introduced a flow-based mathematical formulation and generated three sets of instances for the \ac{2E-VRP} with a maximum of 50 customers and four satellites, based on \ac{vrp} instances. Their branch-and-cut approach is able to solve instances with up to 21 customers to optimality. \cite{perboli2010new} solved additional instances and reduced the optimality gap on others by means of new cutting rules. 

 \cite{Crainic2011b} solved the \ac{2E-VRP} with a multi-start heuristic. The method first assigns customers to satellites heuristically, and then solves the remaining \acp{vrp} with an exact method. In a perturbation step, the assignment of customers to satellites is changed, then the problem is solved again, until a number of iterations is reached.
 
\cite{Jepsen2012} presented a branch-and-cut method, solving 47 out of 93 test instances to optimality, 34 of them for the first time. The authors have been the first to consider a constraint on the number of vehicles \textit{per satellite}, although it was already specified before in the existing data set. This additional constraint had not been taken into account by previous publications. 

\cite{Hemmelmayr2012} developed a metaheuristic based on \ac{ALNS} with a variety of twelve destroy and repair operators. This approach tends to privilege accuracy (high quality solutions) over simplicity and flexibility \citep{Cordeau2002}. The authors also introduced new larger test instances with up to 200 customers and five to ten satellites. Note that the results of ALNS on the problem instances with 50 customers cannot be compared with the proven optimal solutions by \cite{Jepsen2012}, since the algorithm does not consider a limit on the number of vehicles per satellite, but rather a constraint on the total number of vehicles. For most problem instances, this algorithm found the current best known solution or improved it.

\cite{Santos2014} implemented a branch-and-cut-and-price algorithm, which relies on a reformulation of the problem to overcome symmetry issues. They also introduced several classes of valid inequalities. The algorithm performs well in comparison to other exact methods, and they reported solutions for instances with up to 50 customers.

\cite{Baldacci2013} presented a promising exact method to solve the \ac{2E-VRP}. They decomposed the problem into a limited set of \acp{MD-VRP} with side constraints. Detailed results and comparisons with previous publications were provided, as they considered both variants on the instances with 50 customers: with and without the constraint on vehicles per satellite. They also introduced a new set of instances with up to 100 customers.

Recently \cite{Zeng2014a} published a greedy randomized adaptive search procedure, combined with a route-first cluster-second splitting algorithm and a variable neighbourhood descent. They also provide a mathematical formulation. They provide quite good results, although unfortunately their algorithm was tested only with instances comprising up to 50 customers.


\paragraph{Two-echelon location routing problem}
The capacitated \ac{2E-LRP} is by far the most studied version of the \ac{2E-LRP}. Many papers consider location decisions only at the second stage, either because the use of depots is an outcome of the first level routing optimisation, or they consider problems with only one single depot location which is known a priori (\ac{2E-LRPSD}).

\cite{Laporte1988} presented a general analysis of location routing problems and multi-layered problem variants. They compared several mathematical formulations and their computational performance. In a slightly different context, \cite{LaporteNobert1988} formulated a vehicle flow model for the \ac{2E-LRP}. The locations of the depots are assumed to be fixed and unchangeable, such that the location decisions only occur for the satellites. Following the notation by \cite{boccia2011location} they analysed $3/R/\overline{R}$, $3/R/\overline{T}$, $3/T/\overline{R}$, and $3/T/\overline{T}$ problem settings.

\cite{boccia2011location} provided three mathematical formulations for the \ac{2E-LRP} using one-, two-, and three indexed variables inspired from \ac{vrp} and \ac{MD-VRP} formulations. A commercial solver was used to solve some instances generated by the authors, comparing two of the formulations in terms of speed and quality.

\cite{Nguyen2012a} introduced two new sets of instances for the \ac{2E-LRPSD}. They implemented a GRASP with path relinking and a learning process and provided detailed results. In \cite{Nguyen2012b} the authors improved their findings on the same instances by using a multi-start iterated local search. 

\cite{Contardo2012} proposed a branch-and-cut algorithm, which is based on a two-indexed vehicle flow formulation, as well as an \ac{ALNS} heuristic. Both solution approaches were applied to one set of \ac{2E-LRP} instances, and two sets of \ac{2E-LRPSD} instances, outperforming previous heuristics. 

\cite{Schwengerer2012} extended a \ac{VNS} solution approach for the location routing problem from \cite{Pirkwieser2010} and applied it to several instance sets, including the two aforementioned ones with a single depot.

\bigskip

For further details on both problem classes, we refer to the recent survey by \cite{Cuda2014a}. The previous literature review shows that, on the side of exact methods, the best approaches still cannot consistently solve (in practicable time) \ac{2E-VRP} instances with more than 50 customers to optimality. On the side of heuristics, only few methods have been designed and tested to deal with instances with more than 50 delivery locations. To the best of our knowledge, only one heuristic has reported, to this date, computational results on larger \ac{2E-VRP} instances \citep{Hemmelmayr2012}. Hence, there is a need for a more fine-grained study of solution methods for larger problems, as well as for simpler approaches able to efficiently deal with the two families of decisions related to routing and intermediate facilities selection. The proposed method has been designed to cope with these challenges. We developed a technique which performs very well on the classic benchmark instances and, in the meantime, uses fewer and simpler neighbourhood structures than previously published algorithms.
During our research, we finally found inconsistencies regarding different benchmark instances used in previous papers. Some slightly different instances have also been referenced with the same name. Thus we collected the different versions and make them available online with unique names in a uniform file format, as described in Section \ref{instances}.
\section{Mathematical model}
\label{mathmodel}
Different mathematical formulations have been proposed for the \ac{2E-VRP} \citep{Perboli2011, Jepsen2012, Baldacci2013,Santos2013, Santos2014} and for the \ac{2E-LRP} \citep{boccia2011location,Contardo2012}. In this section, we display compact formulations based on the model of \cite{Cuda2014a}.

The \ac{2E-VRP} can be defined on a weighted undirected graph $G=(N,E)$, where the set of vertices $N$ consists of the depot $\{0\}$, the set of possible satellite locations $S=\{1,\dots,|S|\}$ and the set of customers $C=\{|S|+1,\dots,|S|+|C|\}$.
The set of edges $E$ is divided into two subsets, representing the first and second echelon respectively. Set $E^1=\{(i,j):i<j,\; i,j\in\{0\}\cup S\}$ represents the edges which can be traversed by first-level vehicles: those connecting the depot to the satellites, and those interconnecting satellites with each other. The set of edges $E^2=\{(i,j):i<j,\; i,j\in S\cup C, \; (i,j) \notin S \times S\}$ is used for the second level, and corresponds to possible trips between satellite and customers or pairs of customers.
	
A fleet of $v^1$ homogeneous trucks with capacity $Q^1$ is located at the depot. A total of $v^2$ homogeneous city freighters are available, each with a given capacity of $Q^2$. They can be located at any satellite $s \in S$. Still, the number of city freighters at one satellite is limited to $v^2_s$. 

The set $R^1$ contains all possible routes starting from the depot and delivering a given sequence of customers, then returning to the depot again. Similarly each route $r$ in the set of secondary routes $R^2$ starts at a satellite $s \in S$, visits one or several customers in $C$, and returns again to satellite $s$.
Each customer $c \in C$ has a demand of $d_c$ units. Each unit of freight shipped through a satellite induces a handling cost $h_s$.

Given a secondary route $r\in R^2$ and a customer $c\in C$, the parameter $\beta_{rc}\in \{0,1\}$ is equal to 1 if and only if customer $c$ is visited in route $r$, and 0 otherwise. Let $d_r = \sum_{c\in C:c\in r}{d_c} \leq Q^2$ denote the total demand of customers visited in route $r$, and $p_r$ represents the cost of each route $r\in R^1 \cup R^2$. 
The binary variables $x_r \in \{0,1\}$ with $r \in R^1 \cup R^2$ take the value 1 if and only if route $r$ is in the solution. Finally, each decision variable $q_{rs} \geq 0$ with  $r \in R^1,\ s \in S\cap r$, gives the load on the truck on route $r$ that has to be delivered to satellite $s$.

\begin{align}
	\label{obj}
	\min \smashoperator \sum_{r \in R^1 \cup R^2} p_r x_r + \smashoperator \sum_{s \in S} h_s \smashoperator \sum_{r \in R^1} q_{rs}
\end{align}
s.t.
\begin{align}
	\label{trucks}
	\smashoperator \sum_{r\in R^1} x_r &\leq v^1 \\
	\label{cityfr}
	\smashoperator \sum_{r\in R^2} x_r &\leq v^2 \\
	\label{cfpersat}
	\smashoperator \sum_{r\in R^2:s\in r} x_r &\leq v^2_s &s\in S \\	
	\label{cap1}
	\smashoperator \sum_{s\in S \cap r} q_{rs} &\leq Q^1x_r & r\in R^1 \\
	\label{satflow}
	\smashoperator \sum_{r\in R^1} q_{rs} &= \smashoperator \sum_{r\in R^2:s\in r} d_r x_r & s\in S \\
	\label{cust1}
	\smashoperator \sum_{r\in R^2} \beta_{rc} x_r &=1 & c\in C \\
	\label{xbin}
	x_r &\in \{0,1\} & r\in R^1\cup R^2 \\
	\label{qpos}
	q_{rs} &\geq 0 & r\in R^1, s\in S\cap r
\end{align}

The objective function~\eqref{obj} sums up routing costs for all routes on both levels and handling costs per unit moved through each satellite. Constraints~\eqref{trucks}~and~\eqref{cityfr} set the number of available vehicles for trucks and city freighters, respectively. The number of city freighters per satellite is constrained by Constraint~\eqref{cfpersat}. Constraints~\eqref{cap1} ensure that the maximum capacity of the trucks is not exceeded. Constraints~\eqref{satflow} link the quantities of goods between the first and the second level. They guarantee that the incoming goods equal the outgoing goods at the satellites. As there are no split deliveries allowed on the second level, Constraints~\eqref{cust1} ensure that each customer is visited exactly once. The domains of the decision variables are defined by Constraints~\eqref{xbin} and \eqref{qpos}.


The mathematical model for the \ac{2E-LRPSD} is similar to the previous model, but needs some adjustments. Each satellite $s \in S$ has a given opening cost $f_s$ and a capacity of $k_s$ units of freight. There is an unbounded number of vehicles available at both levels. Therefore, Constraints~(\ref{trucks}) to (\ref{cfpersat}) are not needed. The routing of a vehicle incurs fixed costs of $f^1$ for each truck, and $f^2$ for each used city freighter. Each binary parameter $\alpha_{rs}$ is equal to 1 if and only if satellite $s$ is visited on route $r$, and 0 otherwise. If satellite $s$ is opened in the solution, binary variable $y_s$ takes value 1, and 0 otherwise.

Using Constraints (\ref{cap1}) to (\ref{qpos}), the objective function needs to be changed to $\min  \sum_{r \in R^1} (f^1 + p_r) x_r + \sum_{r \in R^2} (f^2 + p_r) x_r + \sum_{s \in S} f_s y_s$ to consider fixed and mileage-based vehicle costs for both levels separately, as well as opening costs for satellites. The capacity limit at the satellites is imposed by \mbox{$\sum_{r \in R^1} q_{rs} \leq k_s y_s$} for all \mbox{$s \in S$}. If a satellite $s$ has been selected to be open, then the delivery by exactly one truck is guaranteed by Constraints \mbox{$\sum_{r \in R^1} \alpha_{rs} x_r = y_s$} for all \mbox{$s \in S$}.

%
\section{Solution method}
\label{method}
The proposed metaheuristic follows the basic structure of a \ac{LNS}, which was first introduced by \cite{shaw1998using}. An initial feasible solution is iteratively destroyed and repaired in order to gradually improve the solution. Such a ruin and recreate approach \citep{Schrimpf2000} has been successfully applied to multiple variants of vehicle routing problems in the past (see, \acs{eg}, \citealt{pisinger2010large}). The destruction of parts of a previous solution (ruin) gives freedom to create a new and better solution (recreate). Algorithm~\ref{a_pseudoLNS} shows the basic structure of the proposed method.

\begin{algorithm}
\DontPrintSemicolon
	\caption{LNS-2E}
	\label{a_pseudoLNS}
	$\mathcal{S}^{best} \leftarrow \mathcal{S} \leftarrow localSearch(repair(instance))$ \label{Alg:init} \tcc*{initial solution}
	$g \leftarrow 0$\;
	\Repeat{$time > time_{max}$}{
		\For{$i \leftarrow 0$ \KwTo $i_{max}$}{
			$\mathcal{S}^{temp} \leftarrow localSearch(repair(destroy(\mathcal{S},g)))$\; \label{Alg:destroy}
				\If{Satellite was opened/closed during previous destroy phase}{
					$g \leftarrow 0$ \label{Alg:grace}																					\tcc*{reset grace period}
				}
			\If{$cost(\mathcal{S}^{temp}) < cost(\mathcal{S})$}{
				$\mathcal{S} \leftarrow \mathcal{S}^{temp}$ 	\label{Alg:newIncumb}						\tcc*{accept better solution}
				$i \leftarrow 0$ \tcc*{reset re-start period}
			}
			$g \leftarrow g+1$\;
		}
		\If{$cost(\mathcal{S}) < cost(\mathcal{S}^{best})$}{
			$\mathcal{S}^{best} \leftarrow \mathcal{S}$ 																		\tcc*{store best solution}
		}
		\Else{
		 $\mathcal{S} \leftarrow localSearch(repair(instance))$  \label{Alg:restart}			\tcc*{re-start: new solution}
		}
	} 
	\Return $\mathcal{S}^{best}$\;
\end{algorithm}

At each iteration of the proposed method, 1) a partial solution destruction is performed on the routes of the second level; 2) then the second level is repaired and improved by means of local search, and finally 3) the first level is reconstructed with a simple heuristic. As such, the first level is constructed from scratch in every iteration, but since the number of nodes in the first-level sub-problem is relatively small, this simple heuristic already finds an optimal or near-optimal solution.

Each of the \emph{destroy} phases performs all the destroy operators sequentially as they are described in Section~\ref{destroyop}. One single \emph{repair} mechanism is used for solution reconstruction and also to obtain an initial solution (Line~\ref{Alg:init}~and~\ref{Alg:destroy}). This procedure is described in Section~\ref{repair}. Afterwards, as needed quantities at the satellites are known, the first level is reconstructed as described in Section~\ref{firstlevel}.

The choices of intermediate facilities may change as a consequence of the repair operator, or through dedicated destroy operators which temporarily close or re-open some possible locations for intermediate facilities. If a change in open or closed satellites has recently taken place, the status of another satellite will not be changed (Line \ref{Alg:destroy}) for a number of iterations that we call \emph{grace period} ($g$~is reset to 0 in Line~\ref{Alg:grace}).

We then put emphasis on a strong local search phase, exploiting well-known procedures like 2-opt \citep{croes1958method}, 2-opt*, or simple relocate and swap moves. \emph{Relocate} shifts a node before or after one of the closest neighbours, if costs are improved. \emph{Swap} explores the exchange of one node with one of the neighbour nodes, as well as exchanging one node with two successive neighbour nodes. This local search phase is applied after the destroy and repair operators. In order to reduce the complexity, moves are only attempted between close customers, as done in the granular search by \cite{Toth2003}. Further information on these moves can be found in the survey by \cite{Vidal2013}.

If a better solution is obtained, it is accepted as the new incumbent solution (Line \ref{Alg:newIncumb}). If no improvement can be found for a large number of iterations, then the algorithm will restart from a new initial solution, even if the objective value is worse (Line \ref{Alg:restart}).

In general, our algorithm requires less local and large neighbourhood operators than the \ac{ALNS} proposed by \cite{Hemmelmayr2012}. The destruction operator parameters are also selected randomly, since the method performed equally well, during our computational experiments, without need for a more advanced adaptive scoring system. In the following, we describe the sets of destroy and repair operators, as well as the management of the decisions related to the first level.

\subsection{Destroy operators}
\label{destroyop}
Our algorithm relies on different destroy operators which are all invoked at each iteration in sequential order. They are applied only to the second level. All of them, except the \textit{open all satellites} neighbourhood, select nodes which are removed from the current solution.

The first four destroy operators are used at each iteration. The last two ones, which change the status of a satellite to \emph{closed} or \emph{open} again, are only invoked if $g$ in Line~\ref{Alg:destroy} of Algorithm~\ref{a_pseudoLNS} has exceeded the grace period $g_{max}$ (i.e. no change in open/closed satellites has taken place recently).
The destroy operators are now described, in their order of use. When applicable, all random samples are uniformly distributed within their given interval.

\paragraph{Related node removal}
A seed customer is randomly chosen. A random number of its Euclidean closest customers as well as the seed customer are removed from the current solution and added to the list of nodes to re-insert. This operator receives a parameter $p_1$, which denotes the maximum percentage of nodes to remove. At most $\lceil p_1 \cdot |C| \rceil$ nodes are removed, with $|C|$ being the overall number of customers.

\paragraph{Biased node removal}
First, the removal cost of each customer is computed: the savings associated to a removal of node $j$, located between $i$ and $k$, is given by $\delta_j = c_{ik}-c_{ij}-c_{jk}$, where $c_{ij}$ denotes the travel cost from node $i$ to node $j$. The probability of selection of a node for removal is then linearly correlated with the delta evaluation value. The higher the gain after removal, the more likely it will be selected and removed. 
In every destroy phase, a random percentage of customers from the interval $[0,p_2]$ is removed.

\paragraph{Random route removal}
Randomly selects routes and removes all containing customers, adding them to the list of nodes to re-insert. This operator randomly selects a number of routes in the interval $[0,\lceil p_3 \cdot \smashoperator\sum_{c \in C} \frac{d_c}{Q^2} \rceil]$.

\paragraph{Remove single node routes}
This operator removes all routes which contain only one single customer. In the case of the \ac{2E-VRP} there is a limited number of overall vehicles available, and thus removing the short routes allows to use a vehicle originating from a different satellite in the next repair phase. This operator is used with a probability of $\hat{p}_4$.
\vspace{\baselineskip}

The last two destroy operators can be used at most at each $g_{max}$ iterations. During this grace period after satellite selection has been actively altered, none of these two operators will be executed.

\paragraph{Close satellite}
Chooses a random satellite. If the satellite can be closed and the remaining open ones still can provide sufficient capacity for a feasible solution, the chosen satellite is closed temporarily. All the customers, which are assigned to it, are removed and added to the list of nodes to re-insert. The satellite stays closed until it is opened again in a later phase. This operation is chosen with a probability of $\hat{p}_5$, given variable $g$ has already exceeded the grace period. If this operator has been executed, $g$ is reset to 0.

\paragraph{Open all satellites}
This neighbourhood makes all previously closed satellites available again. It comes into effect with a probability $\frac {\hat{p}_5}{|S|}$, and thus it depends on the same parameter as the \textit{close satellite} operator and the number of satellites. This operator can only be executed if $g>g_{max}$, \acs{ie} outside the grace period. Its execution resets $g$ to 0.

\subsection{Repair operator, randomisation and initial solution}
\label{repair}
At each repair phase, the insertion of the nodes is done in random order. This repair mechanism can sometimes fail, if one customer remains with a higher demand than the largest free capacity available on any vehicle. In this exceptional case, the repair process is restarted, and the nodes are inserted by decreasing demands to preserve feasibility.

Repair is achieved with a simplified cheapest insertion heuristic. All nodes are sequentially inserted at their cheapest possible position in the solution. The main difference with the classic cheapest insertion heuristic is that the method does not aim to insert the node with the lowest increase in total costs, but just takes the next candidate from the list and inserts it, in order to reduce complexity and enhance solution diversity. It is a simple and greedy heuristic.

Both the initial solution and every partial solution are always repaired by the same operator. The initial solution can be seen as a ``completely destroyed'' solution.

After the second level has been repaired, the local search procedure is performed: 2-opt on each of the routes, 2-opt* on all routes originating at the same satellite. The algorithm then tries to relocate single nodes, swap one node with another and to swap two nodes with one other, within a limited neighbourhood of the $\tau$ closest nodes, again accepting only improvements. This procedure stops when no improving move exists in the entire neighbourhood.
After this procedure, the delivery quantity of each of the satellites is known, and the first level can be constructed using the same insertion heuristic as for the second level and performing local search.

\subsection{Reconstruction of the first level}
\label{firstlevel}
For the \ac{2E-VRP}, it is essential to allow satellites to be delivered by several trucks. In particular, if the demanded quantity at a satellite is larger than a full truckload and no other satellite is available, then there would be no feasible solution. To reconstruct a first level solution, we propose a very simple preprocessing step. Any satellite with a demand larger than a full truckload is virtually duplicated into nodes with demands equal to a truckload, until the remaining demand is smaller than $Q^1$. The same insertion procedure as the repair operator is used to generate a first level solution. This creates back-and-forth trips to the virtual nodes with demands equal to a full truckload, and completes the solution analogously for the remaining nodes.

Usually there are few nodes associated with the \ac{SD-VRP} on the first level. The largest benchmark instances from literature so far contain only ten satellites at most. This very simple policy enabled to find nearly-optimal first level solutions for most considered instances with limited computational effort. Finally, note that in the considered \ac{2E-LRPSD} instances, the capacity of a satellite is never larger than the trucks' capacity. Therefore, split deliveries are not generated during reconstruction.

\section{Computational Experiments}
\label{experiments}
This section describes the currently available sets of instances for the \ac{2E-VRP} (in Section~\ref{instances}) and the used instances for the \ac{2E-LRPSD} (Section \ref{LRPinstances}) and attempts to resolve some inconsistencies. The calibration of the method is described in Section~\ref{sec:parameters}. 
The computational results and the comparisons with other state-of-the-art algorithms are discussed in Sections~\ref{results}~and~\ref{4-38}. Finally, Section~\ref{sec:sensitivity} analyses the sensitivity of the method with respect to several key parameters and design choices.

\subsection{Benchmark Instances for the \acl{2E-VRP}}
\label{instances}
When looking at the literature, it may appear that there are six unique sets of benchmark instances. However, due to inconsistencies with respect to constraints, nomenclature or locations, we identified in fact several different subsets. In what follows, we explain the differences and provide high quality solutions for them.
We consider five different sets of benchmark instances from literature. Sets~2 and 3 were proposed by \cite{Perboli2011} and have been generated based on the instances for the \ac{CVRP} by Christofides and Eilon. Different customers were chosen and converted into satellites. They also proposed the small Set~1 instances, with just twelve customers and two satellites, which we did not consider. Set~4 was proposed by \cite{crainic2010two}; all of them were downloaded from OR-Library \citep{orlib}. 

The instance Sets~2 to 5 as used in \cite{Hemmelmayr2012} were also communicated to us by email \citep{hemmelmayr2013email}. We noticed a few key differences with the ones available from \cite{orlib}.

Set~6 instances were provided from the authors \citep{baldacci2013email}.

All distances are Euclidean, and computed with double precision. Note that handling costs are set to 0 for all sets except 6b. We will now explain the characteristics of these instance sets in detail, and propose unique names for the sets to overcome existing inconsistencies:

\paragraph{Set~2} There are two different versions in circulation: Please note that the instances with 50 customers in the OR-Library contain a mistake\footnote{\label{footnote_capacities}First level vehicles have a capacity of $Q^1 = 160$ units, and second level vehicles, which by design are supposed to be smaller than level 1 trucks, have a capacity of $Q^2 = 400$ units. For instances with two satellites for example, there are 3 trucks available. They can ship a maximum of $3 * 160 = 480$ units. Overall customers' demand $\sum_{c\in C} d_c$ is larger than 480 units, so there is per se no feasible solution for those instances.}. This can be resolved by exchanging $Q^1$ and $Q^2$ capacity values, which is also the way we treated them, like previous authors did.

The names of the instances downloaded from \cite{orlib} and used by \cite{Hemmelmayr2012} were the same, but the instances with 50 customers included different locations for the satellites. For future reference we provide both versions, and we rename the instances with less than 50 customers to Set 2a, the \cite{hemmelmayr2013email} version of 50 customer Set~2 instance files to Set~2b, and the OR-Library version will be called Set~2c. Table \ref{tab:set2} shows the characteristics of all Set~2 instances. 

Instance names used by \cite{Baldacci2013} have the satellite numbers incremented by one. Apart from that they are identical with what we received from \cite{hemmelmayr2013email}. For example Set~2a instance named E-n51-k5-s\textbf{2}-\textbf{17} (Satellites 2 and 17) corresponds to E-n51-k5-s\textbf{3}-\textbf{18} in the result tables of \cite{Baldacci2013}.

We provide both versions (OR-Library with corrected capacities as well as the ones received by Hemmelmayr) with distinguishable names at \url{https://www.univie.ac.at/prolog/research/TwoEVRP}.

\paragraph{Set~3} There are also two different versions of the Set~3 instances in circulation. We collected and solved all of them and distinguished between different versions and identified inconsistencies. Again, the sources \cite{orlib} and \cite{hemmelmayr2013email} were identical for instances with 21 and 32 customers, but different for instances with 50 customers. In the case of Set~3, the filenames for the different instances were also different, so there is no need to introduce new distinguishable names.

The only difference between Set~3 instances with 50 customers from the two sources is the location of the depot. The locations of satellites and customers, as well as the vehicles and demands are identical. All Set~3 instances from \cite{hemmelmayr2013email} place the depot at coordinates (0,0), whereas the files of  \cite{orlib} have the depot located at (30,40). Table \ref{tab:set3depot} shows which instances correspond to each other, apart from satellite location. 

Please also note that like in Set~2, the Set~3 instances with 50 customers from the OR Library also have the capacities of the two vehicle types interchanged (see Footnote \ref{footnote_capacities}). This has been corrected in the files which we provide online.

For easier referencing, we also divide the instances in three parts. Set~3a includes all instances with less than 50 customers, Set~3b the larger instances which have been used by \cite{Hemmelmayr2012}, and Set~3c the larger instances as they are available at the OR-Library, and have been used by \cite{Baldacci2013}, among others.

\paragraph{Set~4} These instances have been treated differently in literature, either with a limit on the number of second level vehicles allowed \textit{per satellite} or only considering a total number of vehicles, with no limitations on the distribution amongst satellites. As proposed in \cite{Baldacci2013}, we solved both versions and follow their nomenclature: Set~4a with the limit per satellite, and Set~4b when the constraint of vehicles per satellite is relaxed.

\paragraph{Set~5} This set of instances has been proposed by \cite{Hemmelmayr2012}. To the best of our knowledge they were the only ones to report solutions on all instances of that set. \cite{Baldacci2013} were able to find solutions on the small instances with only five satellites. 

\paragraph{Set~6} To the best of our knowledge, solutions on these instances have only been reported in \cite{Baldacci2013}. Set~6 includes two subsets: Set~6a, with $h_s = 0$, and Set~6b, which considers different handling costs per freight unit at each of the satellites.
\vspace{\baselineskip}

Table \ref{tab:characteristics} displays an overview of the characteristics of the individual sets. It lists the number of instances in the according set and subset with number of customers (C), satellites (S), trucks (T), city freighters (CF) and available city freighters per satellite $v_s^2$. Column hC shows if the handling costs are non-zero. The source of the instance sets is also provided \citep{hemmelmayr2013email, orlib, baldacci2013email}.

\begin{center}
\begin{threeparttable}[htbp]
	\scriptsize
	\tabcolsep=3pt
  \caption{Characteristics and Sources of Instance Sets}
    \begin{tabular}{rrrrrrrcrccc}
    \toprule
    \multicolumn{1}{c}{Set}   & \multicolumn{1}{c}{Subset} & \multicolumn{1}{c}{Inst.} & \multicolumn{1}{c}{C}     & \multicolumn{1}{c}{S}     & \multicolumn{1}{c}{T}     & \multicolumn{1}{c}{CF}    & \multicolumn{1}{c}{hC} & \multicolumn{1}{c}{$v_s^2$} & \multicolumn{1}{c}{HCC}  & \multicolumn{1}{c}{OR-Library} & \multicolumn{1}{c}{Baldacci} \\
    \midrule
    2     & a     & 6     & 21    & 2     & 3     & 4     & &-     & $\bullet$   & $\bullet$   &  \\
          &       & 6     & 32    & 2     & 3     & 4     & &-     & $\bullet$   & $\bullet$   &  \\
          & b     & 6     & 50    & 2     & 3     & 5     & &-     & $\bullet$   &       &  \\
          &       & 3     & 50    & 4     & 4     & 5     & &-     & $\bullet$   &       &  \\
          & c     & 6     & 50    & 2     & 3     & 5     & &-     &       & $\bullet$   &  \\
          &       & 3     & 50    & 4     & 4     & 5     & &-     &       & $\bullet$   &  \\
					\midrule
    3     & a     & 6     & 21    & 2     & 3     & 4     & &-     & $\bullet$   & $\bullet$   &  \\
          &       & 6     & 32    & 2     & 3     & 4     & &-     & $\bullet$   & $\bullet$   &  \\
          & b     & 6     & 50    & 2     & 3     & 5     & &-     & $\bullet$   &       &  \\
          & c     & 6     & 50    & 2     & 3     & 5     & &-     &       & $\bullet$   &  \\
					\midrule
    4     & a     & 18    & 50    & 2     & 3     & 6     & &4     &       & $\bullet$   &  \\
          &       & 18    & 50    & 3     & 3     & 6     & &3     &       & $\bullet$   &  \\
          &       & 18    & 50    & 5     & 3     & 6     & &2     &       & $\bullet$   &  \\
          & b     & 18    & 50    & 2     & 3     & 6     & &-     & $\bullet$   &       &  \\
          &       & 18    & 50    & 3     & 3     & 6     & &-     & $\bullet$   &       &  \\
          &       & 18    & 50    & 5     & 3     & 6     & &-     & $\bullet$   &       &  \\
					\midrule
    5     &       & 6     & 100   & 5     & 5     & [15,32] & &-     & $\bullet$   &       &  \\
          &       & 6     & 200   & 10    & 5     & [17,35] & &-     & $\bullet$   &       &  \\
          &       & 6     & 200   & 10    & 5     & [30,63] & &-     & $\bullet$   &       &  \\
					\midrule
    6     & a     & 9     & 50    & [4,6] & 2     & 50    & &-     &       &       & $\bullet$ \\
          &       & 9     & 75    & [4,6] & 3     & 75    & &-     &       &       & $\bullet$ \\
          &       & 9     & 100   & [4,6] & 4     & 100   & &-     &       &       & $\bullet$ \\
		      & b     & 9     & 50    & [4,6] & 2     & 50    & $\bullet$ &-     &       &       & $\bullet$ \\
          &       & 9     & 75    & [4,6] & 3     & 75    & $\bullet$ &-     &       &       & $\bullet$ \\
          &       & 9     & 100   & [4,6] & 4     & 100   & $\bullet$ &-     &       &       & $\bullet$ \\
    \bottomrule
    \end{tabular}%
  \label{tab:characteristics}%
\end{threeparttable}%
\end{center}

\subsection{Benchmark Instances for the \acl{2E-LRP}}
\label{LRPinstances}
The proposed algorithm was originally designed for the \ac{2E-VRP}, nevertheless we also tested it on benchmark instances for the \ac{2E-LRPSD}. Two sets, called ``Nguyen'' and ``Prodhon'' are available at \url{http://prodhonc.free.fr/Instances/instances0_us.htm}. They present some small errors or unclear descriptions, which are documented in Appendix \ref{sec:app-LRPinst}. 

\subsection{Parameters}
\label{sec:parameters}
The parameters of the proposed method have been calibrated using meta-calibration: the problem of finding good parameters is assimilated to a black-box optimisation problem, in which the method parameters are the decision variables, and the objective function is simulated by running the method on a set of training instances, containing five randomly selected instances, for each set. To perform a fast optimisation we rely on the \ac{CMA-ES} by \cite{hansen2006}. The source code (in Java) is available at \url{https://www.lri.fr/~hansen/cmaes_inmatlab.html}. 

The performance of our algorithm is rather insensitive to changes in parameters for the small instances, but the rules for closing and opening satellites have to be adjusted to the number of overall available satellites. Our calibration experiments have been conduced for each instance set, independently, and then we searched for one compromise setting for the parameters that yields satisfying results for all different benchmark instances. The calibration results are displayed in Table \ref{tab:parameters}, as well as the average value, standard deviation, and the compromise value which was used for the runs reported in Section \ref{results}.

The size of the limited neighbourhood for the local search relocate and swap moves was also determined by \ac{CMA-ES}. This parameter always converged to $\tau=25$ already in early stages of the tuning process, and thus relocate and swap moves are attempted only for nodes within the radius including the 25 Euclidean closest nodes.

\begin{adjustwidth}{-2cm}{-2cm}
\begin{center}
\begin{threeparttable}[htbp]
	\scriptsize
	\tabcolsep=3pt
  \centering
  \caption{Parameter values obtained by meta-calibration}
    \begin{tabular}{lrrrrrrrrrrrr}
    \toprule
          & \multicolumn{7}{c}{\acs{2E-VRP}}        & \multicolumn{2}{c}{\acs{2E-LRPSD}} &       &       &  \\
    \cmidrule(lr){2-8} \cmidrule(lr){9-10}
          & \multicolumn{1}{c}{Set 2} & \multicolumn{1}{c}{Set 3} & \multicolumn{1}{c}{Set 4a} & \multicolumn{1}{c}{Set 4b} & \multicolumn{1}{c}{Set 5} & \multicolumn{1}{c}{Set 6a} & \multicolumn{1}{c}{Set 6b} & \multicolumn{1}{c}{Nguyen} & \multicolumn{1}{c}{Prodhon} & \multicolumn{1}{c}{Mean}  & \multicolumn{1}{c}{Std. Dev.} & \multicolumn{1}{c}{compromise} \\
					\midrule
    $p_1$  & 0.35  & 0.39  & 0.32  & 0.31  & 0.29  & 0.33  & 0.26  & 0.14  & 0.34  & 0.30  & 0.07  & 0.20 \\
    $p_2$  & 0.18  & 0.20  & 0.12  & 0.07  & 0.80  & 0.52  & 0.78  & 0.50  & 0.19  & 0.37  & 0.27  & 0.35 \\
    $p_3$  & 0.09  & 0.07  & 0.14  & 0.16  & 0.14  & 0.21  & 0.19  & 0.28  & 0.17  & 0.16  & 0.06  & 0.25 \\
    $\hat{p}_4$ & 0.33  & 0.73  & 0.19  & 0.09  & 0.32  & 0.21  & 0.41  & 0.37  & 0.57  & 0.36  & 0.18  & 0.50 \\
    $\hat{p}_5$ & 0.06  & 0.01  & 0.29  & 0.24  & 0.14  & 0.28  & 0.26  & 0.21  & 0.20  & 0.19  & 0.09  & 0.20 \\
    \bottomrule
    \end{tabular}%
  \label{tab:parameters}%
\end{threeparttable}%
\end{center}
\end{adjustwidth}
\subsection{Computational Results}
\label{results}
As done in previous literature, we performed five independent runs on each of the \ac{2E-VRP} benchmark instances and 20 runs on the \ac{2E-LRPSD} instances. The code is written in Java with JDK 1.7.0\_51 and tested on a Intel E5-2670v2 CPU at 2.5 GHz with 3~GB RAM. The code was executed single threaded on one core. We compare the performance of our method on the \ac{2E-VRP} instances with the hybrid GRASP +VND by \cite{Zeng2014a} and the \ac{ALNS} by \cite{Hemmelmayr2012}, when applicable; as well as the currently best known solutions for each instance from the literature. We also show the results of the algorithm on the \ac{2E-LRP} with single depot and compare with the \ac{VNS} of \cite{Schwengerer2012}. We describe the data of the following tables in general and discuss results in detail on each of the instance sets separately.

Tables \ref{tab:set2} to \ref{tab:lrp-p} show the characteristics and detailed results for each instance. The columns \textit{C}, \textit{S}, \textit{T} and \textit{CF} display the main characteristics of the instance, where \textit{C} is the number of customers, \textit{S} is the number of satellites, \textit{T} and \textit{CF} the number of available trucks and city freighters, respectively. The last two columns are not applicable for Tables \ref{tab:lrp-n} to \ref{tab:lrp-p}, as they correspond to \ac{2E-LRPSD} instances with unbounded fleet size.

The next columns display the results of the proposed method (LNS-2E), and methods by \cite{Hemmelmayr2012} (HCC), \cite{Zeng2014a} (ZXXS) for the \ac{2E-VRP} when applicable, and \cite{Schwengerer2012} (SPR) for the \ac{2E-LRPSD}. The average objective value of five runs is given in column \textit{Avg.~5}. Column \textit{Best~5} shows the best solution found within these five runs, and \textit{Best} gives the best objective value found during all experiments, including parameter calibration. Following the work of \cite{Schwengerer2012}, we also used average and best of 20 for the \ac{2E-LRPSD} for better comparison.

Column \textit{t} reports the average overall runtime of the algorithms in seconds, and \textit{t*} the average time when the best solution was found. For easier comparison we chose a simple time limit for termination of our algorithm: 60 seconds for instances with up to 50 customers, and 900 seconds for larger ones. Our time measure corresponds to the wall-clock time of the whole execution of the program, including input and output, computation of the distance matrix, and other pre-processing tasks. \cite{Hemmelmayr2012} and \cite{Schwengerer2012} report CPU times (which may be slightly smaller than wall clock times). \cite{Zeng2014a} only report the time when the best solution was found, but no overall runtime of the algorithm. 

\textit{BKS} refers to the best known solution of that instance. Best known solutions are highlighted in boldface when found by the algorithm, and new BKS are also underlined. We highlight an instance with an asterisk after BKS if the best known solution of the instance is known to be optimal from previous literature.

Tables \ref{tab:set2} and \ref{tab:set3} provide detailed results on the instances of Set~2 and Set~3. HCC, ZXXS and our algorithm find the best known solutions at every run. The solutions have been proven to be optimal for all the instances except Set~2c and Set~3b. To the best of our knowledge, we are the first ones to report solutions on the 2c instances obtained from \cite{orlib}. For Set~3c, the optimal objective values are derived from \cite{Baldacci2013} and \cite{Jepsen2012}, but no results from HCC or ZXXS are available. Summarising Tables~\ref{tab:set2} and \ref{tab:set3}, we can conclude that Sets~2 and 3 are easy in the sense that all runs of all algorithms always found the optimal or best known solution. 

The instances of Set~4 have been addressed in various ways in the literature. \cite{Jepsen2012} considered a limit on the number of city freighters available at each satellite, HCC and ZXXS did not impose this limit, and instead considered the limit on the total number of city freighters only. \cite{Baldacci2013} addressed both variants of the instances to compare their results to both previous results, introducing a new nomenclature: Set~4a for the instances including the limit of vehicles per satellite, and Set~4b when this limit is relaxed.

Tables~\ref{tab:set4a} and \ref{tab:set4b} display the results on Set~4a and 4b instances. 102 out of the 108 instances have been solved to optimality by \cite{Baldacci2013}. Nevertheless we observed small differences of objective values with our solutions (up to a 0.006\% difference). This could be explained by a different rounding convention (we use double precision), or by the small optimality gap of Cplex. As a consequence, bold fonts were used for BKS within 0.006\% precision. In all these cases the underlying solution is identical, just the objective value is marginally different. We can see from Tables~\ref{tab:set4a} and \ref{tab:set4b} that in all cases our best solution corresponds to the optimal or best known solution. Only in 3 and 2 instances of Set~4a and 4b, respectively, some of the runs gave slightly worse solutions. On average, in instance Set~4b our results are slightly better than those of the other heuristics.

To the best of our knowledge, \cite{Hemmelmayr2012} were the only authors who published results on the large Set~5 instances with 10 satellites to this date. \cite{Baldacci2013} report solutions on the small Set~5 instances (100 customers/5 satellites), improving three out of six instances to optimality. The algorithm of \cite{Hemmelmayr2012} was evaluated with a limit of 500 iterations. We compare our results in Table~\ref{tab:set5} and were able to improve the best known solutions on 9 of the 18 instances, depicted with an underlined BKS value. Known optimal solutions are retrieved at least once within the five performed test runs.

Tables \ref{tab:set6a} and \ref{tab:set6b} report the results for the instances of Set~6a and b. From the 54 instances, all except 13 solutions have been proven to be optimal, and on nine of those remaining LNS-2E was able to find better solutions. Best solutions were found typically after less than three minutes.

\begin{adjustwidth}{-2cm}{-2cm}
\begin{center}
\begin{threeparttable}[htbp]
	\scriptsize
	\tabcolsep=3pt
  \caption{Results for Set 2 Instances}
    \begin{tabular}{lccccrrrrrrrrrrr@{\hspace*{0cm}}l}
    \toprule
          &       &       &       &       & \multicolumn{3}{c}{HCC} & \multicolumn{2}{c}{ZXXS} & \multicolumn{5}{c}{LNS-2E}            &       &  \\
					\cmidrule(lr){6-8} \cmidrule(lr){9-10} \cmidrule(lr){11-15}
    \multicolumn{1}{c}{Instance} & \multicolumn{1}{c}{C}     & \multicolumn{1}{c}{S}     & \multicolumn{1}{c}{T}     & \multicolumn{1}{c}{CF}    & \multicolumn{1}{c}{Avg. 5} & \multicolumn{1}{c}{t(s)}  & \multicolumn{1}{c}{t*(s)} & \multicolumn{1}{c}{Avg. 5} & \multicolumn{1}{c}{t*(s)} & \multicolumn{1}{c}{Avg. 5} & \multicolumn{1}{c}{Best 5} & \multicolumn{1}{c}{Best}  & \multicolumn{1}{c}{t(s)}  & \multicolumn{1}{c}{t*(s)} & \multicolumn{1}{c}{\quad BKS}   &  \\
		\midrule
		\multicolumn{17}{l}{\textbf{Set 2a}\tnote{1,2}} \\
    E-n22-k4-s6-17 & 21    & 2     & 3     & 4     & \textbf{417.07} & 37    & 0     & \textbf{417.07} & 0     & \textbf{417.07} & \textbf{417.07} & \textbf{417.07} & 60    & 1     & 417.07 & * \\
    E-n22-k4-s8-14 & 21    & 2     & 3     & 4     & \textbf{384.96} & 34    & 0     & \textbf{384.96} & 0     & \textbf{384.96} & \textbf{384.96} & \textbf{384.96} & 60    & 1     & 384.96 & * \\
    E-n22-k4-s9-19 & 21    & 2     & 3     & 4     & \textbf{470.60} & 35    & 0     & \textbf{470.60} & 0     & \textbf{470.60} & \textbf{470.60} & \textbf{470.60} & 60    & 1     & 470.60 & * \\
    E-n22-k4-s10-14 & 21    & 2     & 3     & 4     & \textbf{371.50} & 37    & 0     & \textbf{371.50} & 0     & \textbf{371.50} & \textbf{371.50} & \textbf{371.50} & 60    & 2     & 371.50 & * \\
    E-n22-k4-s11-12 & 21    & 2     & 3     & 4     & \textbf{427.22} & 31    & 0     & \textbf{427.22} & 0     & \textbf{427.22} & \textbf{427.22} & \textbf{427.22} & 60    & 2     & 427.22 & * \\
    E-n22-k4-s12-16 & 21    & 2     & 3     & 4     & \textbf{392.78} & 36    & 0     & \textbf{392.78} & 0     & \textbf{392.78} & \textbf{392.78} & \textbf{392.78} & 60    & 1     & 392.78 & * \\
    E-n33-k4-s14-22 & 32    & 2     & 3     & 4     & \textbf{779.05} & 85    & 0     & \textbf{730.16} & 0     & \textbf{779.05} & \textbf{779.05} & \textbf{779.05} & 60    & 1     & 779.05 & * \\
    E-n33-k4-s1-9 & 32    & 2     & 3     & 4     & \textbf{730.16} & 74    & 0     & \textbf{714.63} & 0     & \textbf{730.16} & \textbf{730.16} & \textbf{730.16} & 60    & 1     & 730.16 & * \\
    E-n33-k4-s2-13 & 32    & 2     & 3     & 4     & \textbf{714.63} & 64    & 0     & \textbf{707.48} & 0     & \textbf{714.63} & \textbf{714.63} & \textbf{714.63} & 60    & 1     & 714.63 & * \\
    E-n33-k4-s3-17 & 32    & 2     & 3     & 4     & \textbf{707.48} & 58    & 0     & \textbf{778.74} & 1     & \textbf{707.48} & \textbf{707.48} & \textbf{707.48} & 60    & 1     & 707.48 & * \\
    E-n33-k4-s4-5 & 32    & 2     & 3     & 4     & \textbf{778.74} & 77    & 3     & \textbf{756.85} & 0     & \textbf{778.74} & \textbf{778.74} & \textbf{778.74} & 60    & 1     & 778.74 & * \\
    E-n33-k4-s7-25 & 32    & 2     & 3     & 4     & \textbf{756.85} & 53    & 0     & \textbf{779.05} & 0     & \textbf{756.85} & \textbf{756.85} & \textbf{756.85} & 60    & 1     & 756.85 & * \\
		\addlinespace[2pt]
    Avg.  &       &       &       &       & \textbf{577.59} & 51    & 0     & \textbf{577.59} & 0     & \textbf{577.59} & \textbf{577.59} & \textbf{577.59} & 60    & 1     & 577.59 &  \\
		\midrule
    \multicolumn{17}{l}{\textbf{Set 2b}\tnote{1}} \\
    E-n51-k5-s11-19 & 50    & 2     & 3     & 5     & \textbf{581.64} & 182   & 6     & \textbf{597.49} & 1     & \textbf{581.64} & \textbf{581.64} & \textbf{581.64} & 60    & 1     & 581.64 & * \\
    E-n51-k5-s11-19-27-47 & 50    & 4     & 4     & 5     & \textbf{527.63} & 147   & 1     & \textbf{530.76} & 0     & \textbf{527.63} & \textbf{527.63} & \textbf{527.63} & 60    & 4     & 527.63 & * \\
    E-n51-k5-s2-17 & 50    & 2     & 3     & 5     & \textbf{597.49} & 100   & 7     & \textbf{554.81} & 1     & \textbf{597.49} & \textbf{597.49} & \textbf{597.49} & 60    & 3     & 597.49 & * \\
    E-n51-k5-s2-4-17-46 & 50    & 4     & 4     & 5     & \textbf{530.76} & 154   & 1     & \textbf{581.64} & 4     & \textbf{530.76} & \textbf{530.76} & \textbf{530.76} & 60    & 3     & 530.76 & * \\
    E-n51-k5-s27-47 & 50    & 2     & 3     & 5     & \textbf{538.22} & 136   & 1     & \textbf{538.22} & 1     & \textbf{538.22} & \textbf{538.22} & \textbf{538.22} & 60    & 1     & 538.22 & * \\
    E-n51-k5-s32-37 & 50    & 2     & 3     & 5     & \textbf{552.28} & 141   & 1     & \textbf{552.28} & 1     & \textbf{552.28} & \textbf{552.28} & \textbf{552.28} & 60    & 2     & 552.28 & * \\
    E-n51-k5-s4-46 & 50    & 2     & 3     & 5     & \textbf{530.76} & 173   & 0     & \textbf{530.76} & 1     & \textbf{530.76} & \textbf{530.76} & \textbf{530.76} & 60    & 3     & 530.76 & * \\
    E-n51-k5-s6-12 & 50    & 2     & 3     & 5     & \textbf{554.81} & 149   & 2     & \textbf{531.92} & 1     & \textbf{554.81} & \textbf{554.81} & \textbf{554.81} & 60    & 4     & 554.81 & * \\
    E-n51-k5-s6-12-32-37 & 50    & 4     & 4     & 5     & \textbf{531.92} & 150   & 0     & \textbf{527.63} & 1     & \textbf{531.92} & \textbf{531.92} & \textbf{531.92} & 60    & 2     & 531.92 & * \\
		\addlinespace[2pt]
    Avg.  &       &       &       &       & \textbf{549.50} & 148   & 2     & \textbf{549.50} & 1     & \textbf{549.50} & \textbf{549.50} & \textbf{549.50} & 60    & 2     & 549.50 &  \\
		\midrule
    \multicolumn{17}{l}{\textbf{Set 2c}\tnote{2}} \\
    E-n51-k5-s11-19 & 50    & 2     & 3     & 5     &       &       &       &       &       & \textbf{617.42} & \textbf{617.42} & \textbf{617.42} & 60    & 3     & 617.42 &  \\
    E-n51-k5-s11-19-27-47 & 50    & 4     & 4     & 5     &       &       &       &       &       & \textbf{530.76} & \textbf{530.76} & \textbf{530.76} & 60    & 1     & 530.76 &  \\
    E-n51-k5-s2-17 & 50    & 2     & 3     & 5     &       &       &       &       &       & \textbf{601.39} & \textbf{601.39} & \textbf{601.39} & 60    & 3     & 601.39 &  \\
    E-n51-k5-s2-4-17-46 & 50    & 4     & 4     & 5     &       &       &       &       &       & \textbf{601.39} & \textbf{601.39} & \textbf{601.39} & 60    & 4     & 601.39 &  \\
    E-n51-k5-s27-47 & 50    & 2     & 3     & 5     &       &       &       &       &       & \textbf{530.76} & \textbf{530.76} & \textbf{530.76} & 60    & 3     & 530.76 &  \\
    E-n51-k5-s32-37 & 50    & 2     & 3     & 5     &       &       &       &       &       & \textbf{752.59} & \textbf{752.59} & \textbf{752.59} & 60    & 5     & 752.59 &  \\
    E-n51-k5-s4-46 & 50    & 2     & 3     & 5     &       &       &       &       &       & \textbf{702.33} & \textbf{702.33} & \textbf{702.33} & 60    & 4     & 702.33 &  \\
    E-n51-k5-s6-12 & 50    & 2     & 3     & 5     &       &       &       &       &       & \textbf{567.42} & \textbf{567.42} & \textbf{567.42} & 60    & 5     & 567.42 &  \\
    E-n51-k5-s6-12-32-37 & 50    & 4     & 4     & 5     &       &       &       &       &       & \textbf{567.42} & \textbf{567.42} & \textbf{567.42} & 60    & 6     & 567.42 &  \\
		\addlinespace[2pt]
    Avg.  &       &       &       &       &       &       &       &       &       & \textbf{607.94} & \textbf{607.94} & \textbf{607.94} & 60    & 4     & 607.94 &  \\
    \bottomrule
    \end{tabular}%
  \label{tab:set2}%
	
	\begin{tablenotes}\tiny
		\item[1] included in \cite{hemmelmayr2013email}
		\item[2] included in \cite{orlib}
	\end{tablenotes}	
	\quad 
	
\end{threeparttable}%
\end{center}
\end{adjustwidth}

\begin{adjustwidth}{-2cm}{-2cm}
\begin{center}
\begin{threeparttable}[htbp]
	\scriptsize
	\tabcolsep=3pt
  \caption{Results for Set 3 Instances}
    \begin{tabular}{lccccrrrrrrrrrrr@{\hspace*{0cm}}l}
    \toprule
          &       &       &       &       & \multicolumn{3}{c}{HCC} & \multicolumn{2}{c}{ZXXS} & \multicolumn{5}{c}{LNS-2E}            &       &  \\
		\cmidrule(lr){6-8} \cmidrule(lr){9-10} \cmidrule(lr){11-15}			
    \multicolumn{1}{c}{Instance} & \multicolumn{1}{c}{C} & \multicolumn{1}{c}{S} & \multicolumn{1}{c}{T} & \multicolumn{1}{c}{CF} & \multicolumn{1}{c}{Avg. 5} & \multicolumn{1}{c}{t(s)} & \multicolumn{1}{c}{t*(s)} & \multicolumn{1}{c}{Avg. 5} & \multicolumn{1}{c}{t*(s)} & \multicolumn{1}{c}{Avg. 5} & \multicolumn{1}{c}{Best 5} & \multicolumn{1}{c}{Best} & \multicolumn{1}{c}{t(s)} & \multicolumn{1}{c}{t*(s)} & \multicolumn{1}{c}{\quad BKS} &  \\
		\midrule
		\multicolumn{17}{l}{\textbf{Set 3a}\tnote{1,2}} \\
    E-n22-k4-s13-14 & 21    & 2     & 3     & 4     & \textbf{526.15} & 43    & 0     & \textbf{526.15} & 0     & \textbf{526.15} & \textbf{526.15} & \textbf{526.15} & 60    & 2     & 526.15 & * \\
    E-n22-k4-s13-16 & 21    & 2     & 3     & 4     & \textbf{521.09} & 44    & 0     & \textbf{521.09} & 0     & \textbf{521.09} & \textbf{521.09} & \textbf{521.09} & 60    & 2     & 521.09 & * \\
    E-n22-k4-s13-17 & 21    & 2     & 3     & 4     & \textbf{496.38} & 49    & 0     & \textbf{496.38} & 0     & \textbf{496.38} & \textbf{496.38} & \textbf{496.38} & 60    & 1     & 496.38 & * \\
    E-n22-k4-s14-19 & 21    & 2     & 3     & 4     & \textbf{498.80} & 43    & 0     & \textbf{498.80} & 0     & \textbf{498.80} & \textbf{498.80} & \textbf{498.80} & 60    & 1     & 498.80 & * \\
    E-n22-k4-s17-19 & 21    & 2     & 3     & 4     & \textbf{512.81} & 26    & 0     & \textbf{512.81} & 0     & \textbf{512.80} & \textbf{512.80} & \textbf{512.80} & 60    & 4     & 512.80 & * \\
    E-n22-k4-s19-21 & 21    & 2     & 3     & 4     & \textbf{520.42} & 34    & 0     & \textbf{520.42} & 0     & \textbf{520.42} & \textbf{520.42} & \textbf{520.42} & 60    & 2     & 520.42 & * \\
    E-n33-k4-s16-22 & 32    & 2     & 3     & 4     & \textbf{672.17} & 76    & 3     & \textbf{672.17} & 0     & \textbf{672.17} & \textbf{672.17} & \textbf{672.17} & 60    & 4     & 672.17 & * \\
    E-n33-k4-s16-24 & 32    & 2     & 3     & 4     & \textbf{666.02} & 77    & 0     & \textbf{666.02} & 0     & \textbf{666.02} & \textbf{666.02} & \textbf{666.02} & 60    & 1     & 666.02 & * \\
    E-n33-k4-s19-26 & 32    & 2     & 3     & 4     & \textbf{680.36} & 84    & 0     & \textbf{680.36} & 0     & \textbf{680.36} & \textbf{680.36} & \textbf{680.36} & 60    & 1     & 680.36 & * \\
    E-n33-k4-s22-26 & 32    & 2     & 3     & 4     & \textbf{680.37} & 77    & 0     & \textbf{680.37} & 0     & \textbf{680.36} & \textbf{680.36} & \textbf{680.36} & 60    & 1     & 680.36 & * \\
    E-n33-k4-s24-28 & 32    & 2     & 3     & 4     & \textbf{670.43} & 88    & 0     & \textbf{670.43} & 0     & \textbf{670.43} & \textbf{670.43} & \textbf{670.43} & 60    & 2     & 670.43 & * \\
    E-n33-k4-s25-28 & 32    & 2     & 3     & 4     & \textbf{650.58} & 63    & 0     & \textbf{650.58} & 0     & \textbf{650.58} & \textbf{650.58} & \textbf{650.58} & 60    & 1     & 650.58 & * \\
		\addlinespace[2pt]
    Avg.  &       &       &       &       & \textbf{591.30} & 59    & 0     & \textbf{591.30} & 0     & \textbf{591.30} & \textbf{591.30} & \textbf{591.30} & 60    & 2     & 591.30 &  \\
		\midrule
    \multicolumn{17}{l}{\textbf{Set 3b}\tnote{1}} \\
    E-n51-k5-s12-18 & 50    & 2     & 3     & 5     & \textbf{690.59} & 147   & 4     & \textbf{690.59} & 1     & \textbf{690.59} & \textbf{690.59} & \textbf{690.59} & 60    & 8     & 690.59 &  \\
    E-n51-k5-s12-41 & 50    & 2     & 3     & 5     & \textbf{683.05} & 133   & 38    & \textbf{683.05} & 1     & \textbf{683.05} & \textbf{683.05} & \textbf{683.05} & 60    & 11    & 683.05 &  \\
    E-n51-k5-s12-43 & 50    & 2     & 3     & 5     & \textbf{710.41} & 217   & 1     & \textbf{710.41} & 1     & \textbf{710.41} & \textbf{710.41} & \textbf{710.41} & 60    & 5     & 710.41 &  \\
    E-n51-k5-s39-41 & 50    & 2     & 3     & 5     & \textbf{728.54} & 155   & 18    & \textbf{728.54} & 4     & \textbf{728.54} & \textbf{728.54} & \textbf{728.54} & 60    & 7     & 728.54 &  \\
    E-n51-k5-s40-41 & 50    & 2     & 3     & 5     & \textbf{723.75} & 154   & 17    & \textbf{723.75} & 3     & \textbf{723.75} & \textbf{723.75} & \textbf{723.75} & 60    & 5     & 723.75 &  \\
    E-n51-k5-s40-43 & 50    & 2     & 3     & 5     & \textbf{752.15} & 158   & 15    & \textbf{752.15} & 9     & \textbf{752.15} & \textbf{752.15} & \textbf{752.15} & 60    & 12    & 752.15 &  \\
		\addlinespace[2pt]
    Avg.  &       &       &       &       & \textbf{714.75} & 161   & 16    & \textbf{714.75} & 3     & \textbf{714.75} & \textbf{714.75} & \textbf{714.75} & 60    & 8     & 714.75 &  \\
		\midrule
    \multicolumn{17}{l}{\textbf{Set 3c}\tnote{2}} \\
    E-n51-k5-s13-19 & 50    & 2     & 3     & 5     &       &       &       &       &       & \textbf{560.73} & \textbf{560.73} & \textbf{560.73} & 60    & 10    & 560.73 & * \\
    E-n51-k5-s13-42 & 50    & 2     & 3     & 5     &       &       &       &       &       & \textbf{564.45} & \textbf{564.45} & \textbf{564.45} & 60    & 3     & 564.45 & * \\
    E-n51-k5-s13-44 & 50    & 2     & 3     & 5     &       &       &       &       &       & \textbf{564.45} & \textbf{564.45} & \textbf{564.45} & 60    & 2     & 564.45 & * \\
    E-n51-k5-s40-42 & 50    & 2     & 3     & 5     &       &       &       &       &       & \textbf{746.31} & \textbf{746.31} & \textbf{746.31} & 60    & 5     & 746.31 & * \\
    E-n51-k5-s41-42 & 50    & 2     & 3     & 5     &       &       &       &       &       & \textbf{771.56} & \textbf{771.56} & \textbf{771.56} & 60    & 15    & 771.56 & * \\
    E-n51-k5-s41-44 & 50    & 2     & 3     & 5     &       &       &       &       &       & \textbf{802.91} & \textbf{802.91} & \textbf{802.91} & 60    & 16    & 802.91 & * \\
		\addlinespace[2pt]
    Avg.  &       &       &       &       &       &       &       &       &       & \textbf{668.40} & \textbf{668.40} & \textbf{668.40} & 60    & 9     & 668.40 &  \\
    \bottomrule
    \end{tabular}%
\label{tab:set3}%
	\begin{tablenotes}\tiny
		\item[1] included in \cite{hemmelmayr2013email}
		\item[2] included in \cite{orlib}
	\end{tablenotes}	
	\quad 
	
\end{threeparttable}%
\end{center}
\end{adjustwidth}

\begin{adjustwidth}{-2cm}{-2cm}
\begin{center}
\begin{threeparttable}[htbp]
	\scriptsize
  \tabcolsep=3pt
  \caption{Results for Set 4a Instances (with constraint on the number of city freighters per satellite)}
        \begin{tabular}{lccccrrrrrr@{\hspace*{0cm}}l}
    \toprule
    &       &       &       &       & \multicolumn{5}{c}{LNS-2E}               &  &\\
		\cmidrule(lr){6-10}
    \multicolumn{1}{c}{Inst.} & \multicolumn{1}{c}{C}     & \multicolumn{1}{c}{S}     & \multicolumn{1}{c}{T}     & \multicolumn{1}{c}{CF}    & \multicolumn{1}{c}{Avg. 5} & \multicolumn{1}{c}{Best 5} & \multicolumn{1}{c}{Best}  & \multicolumn{1}{c}{t(s)}  & t*(s) & \multicolumn{1}{c}{\quad BKS} &\\
		\midrule
		\multicolumn{12}{l}{\textbf{Set 4a}} \\
    1     & 50    & 2     & 3     & 6     & \textbf{1569.42} & \textbf{1569.42} & \textbf{1569.42} & 60    & 4     & 1569.42 & * \\
    2     & 50    & 2     & 3     & 6     & \textbf{1438.32} & \textbf{1438.32} & \textbf{1438.32} & 60    & 16    & 1438.33 & * \\
    3     & 50    & 2     & 3     & 6     & \textbf{1570.43} & \textbf{1570.43} & \textbf{1570.43} & 60    & 8     & 1570.43 & * \\
    4     & 50    & 2     & 3     & 6     & \textbf{1424.04} & \textbf{1424.04} & \textbf{1424.04} & 60    & 7     & 1424.04 & * \\
    5     & 50    & 2     & 3     & 6     & \textbf{2193.52} & \textbf{2193.52} & \textbf{2193.52} & 60    & 10    & 2193.52 & * \\
    6     & 50    & 2     & 3     & 6     & \textbf{1279.89} & \textbf{1279.89} & \textbf{1279.89} & 60    & 0     & 1279.87 & * \\
    7     & 50    & 2     & 3     & 6     & \textbf{1458.60} & \textbf{1458.60} & \textbf{1458.60} & 60    & 2     & 1458.63 & * \\
    8     & 50    & 2     & 3     & 6     & \textbf{1363.76} & \textbf{1363.76} & \textbf{1363.76} & 60    & 29    & 1363.74 & * \\
    9     & 50    & 2     & 3     & 6     & \textbf{1450.25} & \textbf{1450.25} & \textbf{1450.25} & 60    & 5     & 1450.27 & * \\
    10    & 50    & 2     & 3     & 6     & \textbf{1407.65} & \textbf{1407.65} & \textbf{1407.65} & 60    & 6     & 1407.64 & * \\
    11    & 50    & 2     & 3     & 6     & 2052.21 & \textbf{2047.43} & \textbf{2047.43} & 60    & 3     & 2047.46 & * \\
    12    & 50    & 2     & 3     & 6     & \textbf{1209.46} & \textbf{1209.46} & \textbf{1209.46} & 60    & 8     & 1209.42 & * \\
    13    & 50    & 2     & 3     & 6     & \textbf{1481.80} & \textbf{1481.80} & \textbf{1481.80} & 60    & 7     & 1481.83 & * \\
    14    & 50    & 2     & 3     & 6     & \textbf{1393.64} & \textbf{1393.64} & \textbf{1393.64} & 60    & 1     & 1393.61 & * \\
    15    & 50    & 2     & 3     & 6     & \textbf{1489.92} & \textbf{1489.92} & \textbf{1489.92} & 60    & 16    & 1489.94 & * \\
    16    & 50    & 2     & 3     & 6     & \textbf{1389.20} & \textbf{1389.20} & \textbf{1389.20} & 60    & 2     & 1389.17 & * \\
    17    & 50    & 2     & 3     & 6     & \textbf{2088.48} & \textbf{2088.48} & \textbf{2088.48} & 60    & 15    & 2088.49 & * \\
    18    & 50    & 2     & 3     & 6     & \textbf{1227.68} & \textbf{1227.68} & \textbf{1227.68} & 60    & 1     & 1227.61 & * \\
    19    & 50    & 3     & 3     & 6     & \textbf{1564.66} & \textbf{1564.66} & \textbf{1564.66} & 60    & 3     & 1564.66 & * \\
    20    & 50    & 3     & 3     & 6     & \textbf{1272.98} & \textbf{1272.98} & \textbf{1272.98} & 60    & 25    & 1272.97 & * \\
    21    & 50    & 3     & 3     & 6     & \textbf{1577.82} & \textbf{1577.82} & \textbf{1577.82} & 60    & 2     & 1577.82 & * \\
    22    & 50    & 3     & 3     & 6     & \textbf{1281.83} & \textbf{1281.83} & \textbf{1281.83} & 60    & 3     & 1281.83 & * \\
    23    & 50    & 3     & 3     & 6     & \textbf{1807.35} & \textbf{1807.35} & \textbf{1807.35} & 60    & 8     & 1807.35 & * \\
    24    & 50    & 3     & 3     & 6     & \textbf{1282.69} & \textbf{1282.69} & \textbf{1282.69} & 60    & 0     & 1282.68 & * \\
    25    & 50    & 3     & 3     & 6     & \textbf{1522.40} & \textbf{1522.40} & \textbf{1522.40} & 60    & 4     & 1522.42 & * \\
    26    & 50    & 3     & 3     & 6     & \textbf{1167.47} & \textbf{1167.47} & \textbf{1167.47} & 60    & 1     & 1167.46 & * \\
    27    & 50    & 3     & 3     & 6     & \textbf{1481.56} & \textbf{1481.56} & \textbf{1481.56} & 60    & 42    & 1481.57 & * \\
    28    & 50    & 3     & 3     & 6     & \textbf{1210.46} & \textbf{1210.46} & \textbf{1210.46} & 60    & 3     & 1210.44 & * \\
    29    & 50    & 3     & 3     & 6     & 1722.06 & \textbf{1722.00} & \textbf{1722.00} & 60    & 31    & 1722.04 &  \\
    30    & 50    & 3     & 3     & 6     & \textbf{1211.63} & \textbf{1211.63} & \textbf{1211.63} & 60    & 12    & 1211.59 & * \\
    31    & 50    & 3     & 3     & 6     & \textbf{1490.32} & \textbf{1490.32} & \textbf{1490.32} & 60    & 7     & 1490.34 &  \\
    32    & 50    & 3     & 3     & 6     & \textbf{1199.05} & \textbf{1199.05} & \textbf{1199.05} & 60    & 24    & 1199.00  & * \\
    33    & 50    & 3     & 3     & 6     & \textbf{1508.32} & \textbf{1508.32} & \textbf{1508.32} & 60    & 14    & 1508.30 &  \\
    34    & 50    & 3     & 3     & 6     & \textbf{1233.96} & \textbf{1233.96} & \textbf{1233.96} & 60    & 7     & 1233.92 & * \\
    35    & 50    & 3     & 3     & 6     & \textbf{1718.42} & \textbf{1718.42} & \textbf{1718.42} & 60    & 41    & 1718.41 &  \\
    36    & 50    & 3     & 3     & 6     & \textbf{1228.95} & \textbf{1228.95} & \textbf{1228.95} & 60    & 0     & 1228.89 & * \\
    37    & 50    & 5     & 3     & 6     & \textbf{1528.73} & \textbf{1528.73} & \textbf{1528.73} & 60    & 25    & 1528.73 & * \\
    38    & 50    & 5     & 3     & 6     & \textbf{1169.20} & \textbf{1169.20} & \textbf{1169.20} & 60    & 15    & 1169.20 & * \\
    39    & 50    & 5     & 3     & 6     & \textbf{1520.92} & \textbf{1520.92} & \textbf{1520.92} & 60    & 17    & 1520.92 & * \\
    40    & 50    & 5     & 3     & 6     & \textbf{1199.42} & \textbf{1199.42} & \textbf{1199.42} & 60    & 2     & 1199.42 & * \\
    41    & 50    & 5     & 3     & 6     & \textbf{1667.96} & \textbf{1667.96} & \textbf{1667.96} & 60    & 9     & 1667.96 & * \\
    42    & 50    & 5     & 3     & 6     & \textbf{1194.54} & \textbf{1194.54} & \textbf{1194.54} & 60    & 19    & 1194.54 & * \\
    43    & 50    & 5     & 3     & 6     & \textbf{1439.67} & \textbf{1439.67} & \textbf{1439.67} & 60    & 14    & 1439.67 & * \\
    44    & 50    & 5     & 3     & 6     & \textbf{1045.14} & \textbf{1045.14} & \textbf{1045.14} & 60    & 24    & 1045.13 & * \\
    45    & 50    & 5     & 3     & 6     & 1451.48 & \textbf{1450.95} & \textbf{1450.95} & 60    & 3     & 1450.96 & * \\
    46    & 50    & 5     & 3     & 6     & \textbf{1088.79} & \textbf{1088.79} & \textbf{1088.79} & 60    & 2     & 1088.77 & * \\
    47    & 50    & 5     & 3     & 6     & \textbf{1587.29} & \textbf{1587.29} & \textbf{1587.29} & 60    & 15    & 1587.29 & * \\
    48    & 50    & 5     & 3     & 6     & \textbf{1082.21} & \textbf{1082.21} & \textbf{1082.21} & 60    & 23    & 1082.20 & * \\
    49    & 50    & 5     & 3     & 6     & \textbf{1434.88} & \textbf{1434.88} & \textbf{1434.88} & 60    & 14    & 1434.88 & * \\
    50    & 50    & 5     & 3     & 6     & \textbf{1083.16} & \textbf{1083.16} & \textbf{1083.16} & 60    & 23    & 1083.12 & * \\
    51    & 50    & 5     & 3     & 6     & \textbf{1398.03} & \textbf{1398.03} & \textbf{1398.03} & 60    & 21    & 1398.05 & * \\
    52    & 50    & 5     & 3     & 6     & \textbf{1125.69} & \textbf{1125.69} & \textbf{1125.69} & 60    & 6     & 1125.67 & * \\
    53    & 50    & 5     & 3     & 6     & \textbf{1567.79} & \textbf{1567.79} & \textbf{1567.79} & 60    & 21    & 1567.77 & * \\
    54    & 50    & 5     & 3     & 6     & \textbf{1127.66} & \textbf{1127.66} & \textbf{1127.66} & 60    & 15    & 1127.61 & * \\
		\addlinespace[2pt]
    Avg.  &       &       &       &       & 1420.05 & \textbf{1419.95} & \textbf{1419.95} & 60    & 12    & 1419.94 &  \\
    \bottomrule
    \end{tabular}%
  \label{tab:set4a}%
	\quad 
\end{threeparttable}%
\end{center}
\end{adjustwidth}

\begin{adjustwidth}{-2cm}{-2cm}
\begin{center}
\begin{threeparttable}[htbp]
	\scriptsize
	\tabcolsep=2.7pt	
  \caption{Results for Set 4b Instances ($v_s^2 = v^2$)}
    \begin{tabular}{lccccrrrrrrrrrrrrr@{\hspace*{0cm}}l}
    \toprule
          &       &       &       &       & \multicolumn{4}{c}{HCC}       & \multicolumn{3}{c}{ZXXS} & \multicolumn{5}{c}{LNS-2E}            &        \\
					\cmidrule(lr){6-9} \cmidrule(lr){10-12} \cmidrule(lr){13-17}
    \multicolumn{1}{c}{Inst.} & \multicolumn{1}{c}{C} & \multicolumn{1}{c}{S} & \multicolumn{1}{c}{T} & \multicolumn{1}{c}{CF} & \multicolumn{1}{c}{Avg. 5} & \multicolumn{1}{c}{Best} & \multicolumn{1}{c}{t(s)} & \multicolumn{1}{c}{t*(s)} & \multicolumn{1}{c}{Avg. 5} & \multicolumn{1}{c}{Best} & \multicolumn{1}{c}{t*(s)} & \multicolumn{1}{c}{Avg. 5} & \multicolumn{1}{c}{Best 5} & \multicolumn{1}{c}{Best} & \multicolumn{1}{c}{t(s)} & \multicolumn{1}{c}{t*(s)} & \multicolumn{1}{c}{\quad BKS}   \\
		\midrule
    \multicolumn{18}{l}{\textbf{Set 4b}} \\		
    1     & 50    & 2     & 3     & 6     & \textbf{1569.42} & \textbf{1569.42} & 235   & 6     & \textbf{1569.42} & \textbf{1569.42} & 1     & \textbf{1569.42} & \textbf{1569.42} & \textbf{1569.42} & 60    & 15    & 1569.42 & * \\
    2     & 50    & 2     & 3     & 6     & 1441.02 & \textbf{1438.33} & 155   & 43    & \textbf{1438.33} & \textbf{1438.33} & 40    & \textbf{1438.32} & \textbf{1438.32} & \textbf{1438.32} & 60    & 6     & 1438.33 & * \\
    3     & 50    & 2     & 3     & 6     & \textbf{1570.43} & \textbf{1570.43} & 183   & 3     & \textbf{1570.43} & \textbf{1570.43} & 1     & \textbf{1570.43} & \textbf{1570.43} & \textbf{1570.43} & 60    & 7     & 1570.43 & * \\
    4     & 50    & 2     & 3     & 6     & \textbf{1424.04} & \textbf{1424.04} & 130   & 11    & 1429.04 & \textbf{1424.04} & 73    & \textbf{1424.04} & \textbf{1424.04} & \textbf{1424.04} & 60    & 7     & 1424.04 & * \\
    5     & 50    & 2     & 3     & 6     & \textbf{2194.11} & \textbf{2194.11} & 614   & 63    & \textbf{2193.52} & \textbf{2193.52} & 34    & \textbf{2193.52} & \textbf{2193.52} & \textbf{2193.52} & 60    & 18    & 2193.52 & * \\
    6     & 50    & 2     & 3     & 6     & \textbf{1279.87} & \textbf{1279.87} & 99    & 2     & \textbf{1279.87} & \textbf{1279.87} & 1     & \textbf{1279.89} & \textbf{1279.89} & \textbf{1279.89} & 60    & 0     & 1279.87 & * \\
    7     & 50    & 2     & 3     & 6     & 1458.63 & 1458.63 & 169   & 6     & \textbf{1408.57} & \textbf{1408.57} & 16    & \textbf{1408.58} & \textbf{1408.58} & \textbf{1408.58} & 60    & 17    & 1408.57 & * \\
    8     & 50    & 2     & 3     & 6     & \textbf{1360.32} & \textbf{1360.32} & 205   & 5     & \textbf{1360.32} & \textbf{1360.32} & 4     & \textbf{1360.32} & \textbf{1360.32} & \textbf{1360.32} & 60    & 8     & 1360.32 & * \\
    9     & 50    & 2     & 3     & 6     & 1450.27 & 1450.27 & 204   & 46    & \textbf{1403.53} & \textbf{1403.53} & 4     & \textbf{1403.53} & \textbf{1403.53} & \textbf{1403.53} & 60    & 11    & 1403.53 & * \\
    10    & 50    & 2     & 3     & 6     & \textbf{1360.56} & \textbf{1360.56} & 174   & 1     & \textbf{1360.56} & \textbf{1360.56} & 1     & \textbf{1360.54} & \textbf{1360.54} & \textbf{1360.54} & 60    & 1     & 1360.56 & * \\
    11    & 50    & 2     & 3     & 6     & 2059.88 & 2059.88 & 648   & 101   & 2059.41 & 2059.41 & 4     & 2054.60 & \textbf{2047.43} & \textbf{2047.43} & 60    & 2     & 2047.46 & * \\
    12    & 50    & 2     & 3     & 6     & \textbf{1209.42} & \textbf{1209.42} & 205   & 44    & \textbf{1209.42} & \textbf{1209.42} & 6     & \textbf{1209.46} & \textbf{1209.46} & \textbf{1209.46} & 60    & 20    & 1209.42 & * \\
    13    & 50    & 2     & 3     & 6     & 1481.83 & 1481.83 & 220   & 25    & \textbf{1450.93} & \textbf{1450.93} & 2     & \textbf{1450.95} & \textbf{1450.94} & \textbf{1450.94} & 60    & 10    & 1450.93 & * \\
    14    & 50    & 2     & 3     & 6     & \textbf{1393.61} & \textbf{1393.61} & 189   & 6     & \textbf{1393.61} & \textbf{1393.61} & 1     & \textbf{1393.64} & \textbf{1393.64} & \textbf{1393.64} & 60    & 3     & 1393.61 & * \\
    15    & 50    & 2     & 3     & 6     & 1489.94 & 1489.94 & 173   & 9     & \textbf{1466.83} & \textbf{1466.83} & 1     & \textbf{1466.84} & \textbf{1466.84} & \textbf{1466.84} & 60    & 2     & 1466.83 & * \\
    16    & 50    & 2     & 3     & 6     & \textbf{1387.83} & \textbf{1387.83} & 147   & 6     & \textbf{1387.83} & \textbf{1387.83} & 6     & \textbf{1387.85} & \textbf{1387.85} & \textbf{1387.85} & 60    & 12    & 1387.83 & * \\
    17    & 50    & 2     & 3     & 6     & \textbf{2088.49} & \textbf{2088.49} & 625   & 165   & \textbf{2088.49} & \textbf{2088.49} & 27    & \textbf{2088.48} & \textbf{2088.48} & \textbf{2088.48} & 60    & 13    & 2088.49 & * \\
    18    & 50    & 2     & 3     & 6     & \textbf{1227.61} & \textbf{1227.61} & 94    & 3     & \textbf{1227.61} & \textbf{1227.61} & 1     & \textbf{1227.68} & \textbf{1227.68} & \textbf{1227.68} & 60    & 5     & 1227.61 & * \\
    19    & 50    & 3     & 3     & 6     & \textbf{1546.28} & \textbf{1546.28} & 171   & 25    & \textbf{1546.28} & \textbf{1546.28} & 25    & \textbf{1546.28} & \textbf{1546.28} & \textbf{1546.28} & 60    & 34    & 1546.28 & * \\
    20    & 50    & 3     & 3     & 6     & \textbf{1272.97} & \textbf{1272.97} & 99    & 12    & \textbf{1272.97} & \textbf{1272.97} & 57    & \textbf{1272.98} & \textbf{1272.98} & \textbf{1272.98} & 60    & 11    & 1272.97 & * \\
    21    & 50    & 3     & 3     & 6     & \textbf{1577.82} & \textbf{1577.82} & 155   & 61    & \textbf{1577.82} & \textbf{1577.82} & 19    & \textbf{1577.82} & \textbf{1577.82} & \textbf{1577.82} & 60    & 16    & 1577.82 & * \\
    22    & 50    & 3     & 3     & 6     & \textbf{1281.83} & \textbf{1281.83} & 127   & 2     & \textbf{1281.83} & \textbf{1281.83} & 28    & \textbf{1281.83} & \textbf{1281.83} & \textbf{1281.83} & 60    & 4     & 1281.83 & * \\
    23    & 50    & 3     & 3     & 6     & \textbf{1652.98} & \textbf{1652.98} & 175   & 5     & \textbf{1652.98} & \textbf{1652.98} & 3     & \textbf{1652.98} & \textbf{1652.98} & \textbf{1652.98} & 60    & 4     & 1652.98 & * \\
    24    & 50    & 3     & 3     & 6     & \textbf{1282.68} & \textbf{1282.68} & 110   & 2     & \textbf{1282.68} & \textbf{1282.68} & 1     & \textbf{1282.69} & \textbf{1282.69} & \textbf{1282.69} & 60    & 1     & 1282.68 & * \\
    25    & 50    & 3     & 3     & 6     & 1440.84 & 1440.68 & 154   & 53    & \textbf{1408.57} & \textbf{1408.57} & 17    & \textbf{1408.58} & \textbf{1408.58} & \textbf{1408.58} & 60    & 20    & 1408.57 & * \\
    26    & 50    & 3     & 3     & 6     & \textbf{1167.46} & \textbf{1167.46} & 96    & 0     & \textbf{1167.46} & \textbf{1167.46} & 7     & \textbf{1167.47} & \textbf{1167.47} & \textbf{1167.47} & 60    & 16    & 1167.46 & * \\
    27    & 50    & 3     & 3     & 6     & 1447.79 & \textbf{1444.50} & 163   & 12    & 1454.63 & \textbf{1444.51} & 39    & \textbf{1444.49} & \textbf{1444.49} & \textbf{1444.49} & 60    & 20    & 1444.50 & * \\
    28    & 50    & 3     & 3     & 6     & \textbf{1210.44} & \textbf{1210.44} & 143   & 7     & \textbf{1210.44} & \textbf{1210.44} & 2     & \textbf{1210.46} & \textbf{1210.46} & \textbf{1210.46} & 60    & 6     & 1210.44 & * \\
    29    & 50    & 3     & 3     & 6     & 1561.81 & 1559.82 & 178   & 102   & 1555.56 & \textbf{1552.66} & 30    & \textbf{1552.66} & \textbf{1552.66} & \textbf{1552.66} & 60    & 39    & 1552.66 & * \\
    30    & 50    & 3     & 3     & 6     & \textbf{1211.59} & \textbf{1211.59} & 132   & 5     & \textbf{1211.59} & \textbf{1211.59} & 1     & \textbf{1211.63} & \textbf{1211.63} & \textbf{1211.63} & 60    & 5     & 1211.59 & * \\
    31    & 50    & 3     & 3     & 6     & \textbf{1440.86} & \textbf{1440.86} & 144   & 37    & \textbf{1441.07} & \textbf{1440.86} & 29    & \textbf{1440.85} & \textbf{1440.85} & \textbf{1440.85} & 60    & 15    & 1440.86 & * \\
    32    & 50    & 3     & 3     & 6     & \textbf{1199.00} & \textbf{1199.00} & 102   & 11    & \textbf{1199.00} & \textbf{1199.00} & 10    & \textbf{1199.05} & \textbf{1199.05} & \textbf{1199.05} & 60    & 21    & 1199.00 & * \\
    33    & 50    & 3     & 3     & 6     & \textbf{1478.86} & \textbf{1478.86} & 159   & 16    & \textbf{1478.86} & \textbf{1478.86} & 14    & \textbf{1478.87} & \textbf{1478.87} & \textbf{1478.87} & 60    & 15    & 1478.86 & * \\
    34    & 50    & 3     & 3     & 6     & \textbf{1233.92} & \textbf{1233.92} & 93    & 4     & \textbf{1233.92} & \textbf{1233.92} & 11    & \textbf{1233.96} & \textbf{1233.96} & \textbf{1233.96} & 60    & 11    & 1233.92 & * \\
    35    & 50    & 3     & 3     & 6     & \textbf{1570.80} & \textbf{1570.72} & 182   & 116   & \textbf{1570.72} & \textbf{1570.72} & 4     & \textbf{1570.73} & \textbf{1570.73} & \textbf{1570.73} & 60    & 10    & 1570.72 & * \\
    36    & 50    & 3     & 3     & 6     & \textbf{1228.89} & \textbf{1228.89} & 123   & 6     & \textbf{1228.89} & \textbf{1228.89} & 2     & \textbf{1228.95} & \textbf{1228.95} & \textbf{1228.95} & 60    & 6     & 1228.89 & * \\
    37    & 50    & 5     & 3     & 6     & \textbf{1528.81} & \textbf{1528.73} & 143   & 55    & \textbf{1528.98} & \textbf{1528.73} & 43    & \textbf{1528.73} & \textbf{1528.73} & \textbf{1528.73} & 60    & 27    & 1528.73 &  \\
    38    & 50    & 5     & 3     & 6     & \textbf{1163.07} & \textbf{1163.07} & 88    & 15    & \textbf{1163.07} & \textbf{1163.07} & 1     & \textbf{1163.07} & \textbf{1163.07} & \textbf{1163.07} & 60    & 17    & 1163.07 & * \\
    39    & 50    & 5     & 3     & 6     & \textbf{1520.92} & \textbf{1520.92} & 158   & 33    & \textbf{1520.92} & \textbf{1520.92} & 16    & \textbf{1520.92} & \textbf{1520.92} & \textbf{1520.92} & 60    & 25    & 1520.92 & * \\
    40    & 50    & 5     & 3     & 6     & 1165.24 & \textbf{1163.04} & 84    & 20    & \textbf{1163.04} & \textbf{1163.04} & 7     & \textbf{1163.04} & \textbf{1163.04} & \textbf{1163.04} & 60    & 5     & 1163.04 & * \\
    41    & 50    & 5     & 3     & 6     & \textbf{1652.98} & \textbf{1652.98} & 150   & 12    & \textbf{1652.98} & \textbf{1652.98} & 9     & \textbf{1652.98} & \textbf{1652.98} & \textbf{1652.98} & 60    & 14    & 1652.98 & * \\
    42    & 50    & 5     & 3     & 6     & \textbf{1190.17} & \textbf{1190.17} & 95    & 31    & \textbf{1190.17} & \textbf{1190.17} & 71    & \textbf{1190.17} & \textbf{1190.17} & \textbf{1190.17} & 60    & 9     & 1190.17 & * \\
    43    & 50    & 5     & 3     & 6     & 1408.95 & \textbf{1406.11} & 151   & 60    & \textbf{1406.11} & \textbf{1406.11} & 24    & 1407.09 & \textbf{1406.10} & \textbf{1406.10} & 60    & 13    & 1406.11 & * \\
    44    & 50    & 5     & 3     & 6     & \textbf{1035.32} & \textbf{1035.03} & 109   & 30    & \textbf{1035.03} & \textbf{1035.03} & 7     & \textbf{1035.05} & \textbf{1035.05} & \textbf{1035.05} & 60    & 42    & 1035.03 & * \\
    45    & 50    & 5     & 3     & 6     & 1406.43 & 1403.10 & 144   & 104   & \textbf{1402.03} & \textbf{1402.03} & 27    & \textbf{1401.87} & \textbf{1401.87} & \textbf{1401.87} & 60    & 21    & 1401.87 & * \\
    46    & 50    & 5     & 3     & 6     & 1058.97 & \textbf{1058.11} & 74    & 17    & \textbf{1058.11} & \textbf{1058.11} & 7     & \textbf{1058.10} & \textbf{1058.10} & \textbf{1058.10} & 60    & 16    & 1058.11 & * \\
    47    & 50    & 5     & 3     & 6     & 1564.41 & 1559.82 & 185   & 103   & 1557.04 & \textbf{1552.66} & 25    & \textbf{1552.66} & \textbf{1552.66} & \textbf{1552.66} & 60    & 24    & 1552.66 & * \\
    48    & 50    & 5     & 3     & 6     & \textbf{1074.50} & \textbf{1074.50} & 83    & 2     & \textbf{1074.50} & \textbf{1074.50} & 1     & \textbf{1074.51} & \textbf{1074.51} & \textbf{1074.51} & 60    & 4     & 1074.50 & * \\
    49    & 50    & 5     & 3     & 6     & \textbf{1435.28} & \textbf{1434.88} & 140   & 81    & \textbf{1434.88} & \textbf{1434.88} & 38    & \textbf{1434.88} & \textbf{1434.88} & \textbf{1434.88} & 60    & 9     & 1434.88 &  \\
    50    & 50    & 5     & 3     & 6     & \textbf{1065.25} & \textbf{1065.25} & 92    & 16    & \textbf{1065.25} & \textbf{1065.25} & 2     & \textbf{1065.30} & \textbf{1065.30} & \textbf{1065.30} & 60    & 33    & 1065.25 & * \\
    51    & 50    & 5     & 3     & 6     & \textbf{1387.72} & \textbf{1387.51} & 138   & 46    & \textbf{1387.51} & \textbf{1387.51} & 3     & \textbf{1387.51} & \textbf{1387.51} & \textbf{1387.51} & 60    & 19    & 1387.51 & * \\
    52    & 50    & 5     & 3     & 6     & \textbf{1103.76} & \textbf{1103.42} & 102   & 47    & \textbf{1103.42} & \textbf{1103.42} & 22    & \textbf{1103.47} & \textbf{1103.47} & \textbf{1103.47} & 60    & 17    & 1103.42 & * \\
    53    & 50    & 5     & 3     & 6     & \textbf{1545.73} & \textbf{1545.73} & 148   & 37    & \textbf{1545.73} & \textbf{1545.73} & 4     & \textbf{1545.76} & \textbf{1545.76} & \textbf{1545.76} & 60    & 21    & 1545.73 & * \\
    54    & 50    & 5     & 3     & 6     & \textbf{1113.62} & \textbf{1113.62} & 90    & 2     & \textbf{1113.62} & \textbf{1113.62} & 12    & \textbf{1113.66} & \textbf{1113.66} & \textbf{1113.66} & 60    & 5     & 1113.62 & * \\
		\addlinespace[2pt] 		
    Avg.  &       &       &       &       & 1401.39 & 1400.96 & 169   & 32    & 1397.69 & 1397.27 & 16    & 1397.21 & \textbf{1397.06} & \textbf{1397.06} & 60    & 14    & 1397.04 &  \\
    \bottomrule
    \end{tabular}%
  \label{tab:set4b}%
	\quad 
	
\end{threeparttable}%
\end{center}
\end{adjustwidth}
\begin{adjustwidth}{-2cm}{-2cm}
\begin{center}
\begin{threeparttable}[htbp]
	\scriptsize
	\tabcolsep=2.6pt
  \caption{Results for Set 5 Instances}
		\begin{tabular}{lccccrrrrrrrrrr@{\hspace*{0cm}}l}
    \toprule
    &       &       &       &       & \multicolumn{4}{c}{HCC}       & \multicolumn{5}{c}{LNS-2E}               &  &\\
		\cmidrule(lr){6-9} \cmidrule(lr){10-14}
    \multicolumn{1}{c}{Instance} & \multicolumn{1}{c}{C}     & \multicolumn{1}{c}{S}     & \multicolumn{1}{c}{T}     & \multicolumn{1}{c}{CF}    & \multicolumn{1}{c}{Avg. 5} & \multicolumn{1}{c}{Best}  & \multicolumn{1}{c}{t(s)}  & t*(s) & \multicolumn{1}{c}{Avg. 5} & \multicolumn{1}{c}{Best 5} & \multicolumn{1}{c}{Best}  & \multicolumn{1}{c}{t(s)}  & t*(s) & \multicolumn{1}{c}{\quad BKS} &\\
		\midrule
    \multicolumn{16}{l}{\textbf{Set 5}} \\		
    100-5-1 & 100   & 5     & 5     & 32    & 1588.73 & 1565.45 & 353   & 116   & 1566.87 & \textbf{1564.46} & \textbf{1564.46} & 900   & 201   & 1564.46 & * \\
    100-5-1b & 100   & 5     & 5     & 15    & 1126.93 & 1111.34 & 397   & 45    & 1111.93 & \textbf{1108.62} & \textbf{1108.62} & 900   & 176   & \underline{1108.62} &  \\
    100-5-2 & 100   & 5     & 5     & 32    & 1022.29 & \textbf{1016.32} & 406   & 117   & 1017.94 & \textbf{1016.32} & \textbf{1016.32} & 900   & 75    & 1016.32 & * \\
    100-5-2b & 100   & 5     & 5     & 15    & 789.05 & \textbf{782.25} & 340   & 170   & 783.07 & \textbf{782.25} & \textbf{782.25} & 900   & 152   & 782.25 &  \\
    100-5-3 & 100   & 5     & 5     & 30    & 1046.67 & \textbf{1045.29} & 352   & 80    & \textbf{1045.29} & \textbf{1045.29} & \textbf{1045.29} & 900   & 131   & 1045.29 & * \\
    100-5-3b & 100   & 5     & 5     & 16    & 828.99 & 828.99 & 391   & 127   & \textbf{828.54} & \textbf{828.54} & \textbf{828.54} & 900   & 62    & \underline{828.54} &  \\
    100-10-1 & 100   & 10    & 5     & 35    & 1137.00 & 1130.23 & 429   & 136   & 1132.11 & 1125.53 & \textbf{1124.93} & 900   & 567   & \underline{1124.93} &  \\
    100-10-1b & 100   & 10    & 5     & 18    & 928.01 & 916.48 & 476   & 262   & 922.85 & \textbf{916.25} & \textbf{916.25} & 900   & 424   & \underline{916.25} &  \\
    100-10-2 & 100   & 10    & 5     & 33    & 1009.49 & \textbf{990.58} & 356   & 232   & 1014.61 & 1012.14 & 1002.15 & 900   & 471   & 990.58 &  \\
    100-10-2b & 100   & 10    & 5     & 18    & 773.58 & \textbf{768.61} & 432   & 157   & 786.64 & 781.27 & 774.11 & 900   & 416   & 768.61 &  \\
    100-10-3 & 100   & 10    & 5     & 32    & 1055.28 & \textbf{1043.25} & 415   & 209   & 1053.55 & 1049.77 & 1048.53 & 900   & 105   & 1043.25 &  \\
    100-10-3b & 100   & 10    & 5     & 17    & 861.88 & \textbf{850.92} & 418   & 29    & 858.72 & 854.90 & 854.90 & 900   & 175   & 850.92 &  \\
    200-10-1 & 200   & 10    & 5     & 62    & 1626.83 & 1574.12 & 888   & 207   & 1598.46 & 1580.34 & \textbf{1556.79} & 900   & 730   & \underline{1556.79} &  \\
    200-10-1b & 200   & 10    & 5     & 30    & 1239.79 & 1201.75 & 692   & 374   & 1217.23 & 1191.59 & \textbf{1187.62} & 900   & 588   & \underline{1187.62} &  \\
    200-10-2 & 200   & 10    & 5     & 63    & 1416.87 & 1374.74 & 1072  & 496   & 1406.16 & 1366.36 & \textbf{1365.74} & 900   & 534   & \underline{1365.74} &  \\
    200-10-2b & 200   & 10    & 5     & 30    & 1018.57 & 1003.75 & 1058  & 221   & 1016.05 & 1008.46 & \textbf{1002.85} & 900   & 721   & \underline{1002.85} &  \\
    200-10-3 & 200   & 10    & 5     & 63    & 1808.24 & \textbf{1787.73} & 916   & 305   & 1809.44 & 1797.80 & 1793.99 & 900   & 675   & 1787.73 &  \\
    200-10-3b & 200   & 10    & 5     & 30    & 1208.38 & 1200.74 & 1217  & 478   & 1206.85 & 1202.21 & \textbf{1197.90} & 900   & 523   & \underline{1197.90} &  \\
		\addlinespace[2pt]
    Avg.  &       &       &       &       & 1138.14 & 1121.81 & 589   & 209   & 1132.02 & 1124.01 & 1120.62 & 900   & 374   & 1118.81 &  \\
    \bottomrule
    \end{tabular}%
  \label{tab:set5}%
\quad	
\end{threeparttable}%
\end{center}
\end{adjustwidth}

\begin{adjustwidth}{-2cm}{-2cm}
\begin{center}
\begin{threeparttable}[htbp]
	\scriptsize
  \tabcolsep=3pt
  \caption{Results for Set 6a Instances (no handling costs)}
	\begin{tabular}{lccccrrrrrr@{\hspace*{0cm}}l}
    \toprule
          &       &       &       &       & \multicolumn{5}{c}{LNS-2E}               &  &\\
					\cmidrule(lr){6-10}
    \multicolumn{1}{c}{Instance} & \multicolumn{1}{c}{C} & \multicolumn{1}{c}{S} & \multicolumn{1}{c}{T} & \multicolumn{1}{c}{CF} & \multicolumn{1}{c}{Avg. 5} & \multicolumn{1}{c}{Best 5} & \multicolumn{1}{c}{Best} & \multicolumn{1}{c}{t(s)} & t*(s) & \multicolumn{1}{c}{\quad BKS} &\\
		\midrule
    \multicolumn{12}{l}{\textbf{Set 6a}}   \\				
    A-n51-4 & 50    & 4     & 2     & 50    & \textbf{652.00} & \textbf{652.00} & \textbf{652.00} & 60    & 19    & 652.00 & * \\
    A-n51-5 & 50    & 5     & 2     & 50    & \textbf{663.41} & \textbf{663.41} & \textbf{663.41} & 60    & 37    & 663.41 & * \\
    A-n51-6 & 50    & 6     & 2     & 50    & \textbf{662.51} & \textbf{662.51} & \textbf{662.51} & 60    & 23    & 662.51 & * \\
    A-n76-4 & 75    & 4     & 3     & 75    & 985.98 & \textbf{985.95} & \textbf{985.95} & 900   & 128   & 985.95 & * \\
    A-n76-5 & 75    & 5     & 3     & 75    & 981.19 & \textbf{979.15} & \textbf{979.15} & 900   & 286   & 979.15 & * \\
    A-n76-6 & 75    & 6     & 3     & 75    & 971.65 & \textbf{970.20} & \textbf{970.20} & 900   & 233   & 970.20 & * \\
    A-n101-4 & 100   & 4     & 4     & 100   & 1194.38 & \textbf{1194.17} & \textbf{1194.17} & 900   & 267   & 1194.17 & * \\
    A-n101-5 & 100   & 5     & 4     & 100   & 1215.89 & 1211.40 & 1211.40 & 900   & 414   & 1211.38 & * \\
    A-n101-6 & 100   & 6     & 4     & 100   & 1161.91 & 1158.97 & \textbf{1155.96} & 900   & 154   & \underline{1155.96} &  \\
    B-n51-4 & 50    & 4     & 2     & 50    & \textbf{563.98} & \textbf{563.98} & \textbf{563.98} & 60    & 6     & 563.98 & * \\
    B-n51-5 & 50    & 5     & 2     & 50    & \textbf{549.23} & \textbf{549.23} & \textbf{549.23} & 60    & 55    & 549.23 & * \\
    B-n51-6 & 50    & 6     & 2     & 50    & \textbf{556.32} & \textbf{556.32} & \textbf{556.32} & 60    & 39    & 556.32 & * \\
    B-n76-4 & 75    & 4     & 3     & 75    & 793.97 & \textbf{792.73} & \textbf{792.73} & 900   & 320   & 792.73 & * \\
    B-n76-5 & 75    & 5     & 3     & 75    & 784.27 & 784.19 & 784.19 & 900   & 190   & 783.93 & * \\
    B-n76-6 & 75    & 6     & 3     & 75    & 775.75 & 774.24 & \textbf{774.17} & 900   & 160   & 774.17 & * \\
    B-n101-4 & 100   & 4     & 4     & 100   & 939.79 & \textbf{939.21} & \textbf{939.21} & 900   & 377   & 939.21 & * \\
    B-n101-5 & 100   & 5     & 4     & 100   & 971.27 & 969.13 & 969.13 & 900   & 161   & 967.82 & * \\
    B-n101-6 & 100   & 6     & 4     & 100   & 961.91 & \textbf{960.29} & \textbf{960.29} & 900   & 88    & 960.29 & * \\
    C-n51-4 & 50    & 4     & 2     & 50    & \textbf{689.18} & \textbf{689.18} & \textbf{689.18} & 60    & 13    & 689.18 & * \\
    C-n51-5 & 50    & 5     & 2     & 50    & \textbf{723.12} & \textbf{723.12} & \textbf{723.12} & 60    & 19    & 723.12 & * \\
    C-n51-6 & 50    & 6     & 2     & 50    & \textbf{697.00} & \textbf{697.00} & \textbf{697.00} & 60    & 46    & 697.00 & * \\
    C-n76-4 & 75    & 4     & 3     & 75    & 1055.61 & \textbf{1054.89} & \textbf{1054.89} & 900   & 339   & 1054.89 & * \\
    C-n76-5 & 75    & 5     & 3     & 75    & \textbf{1115.32} & \textbf{1115.32} & \textbf{1115.32} & 900   & 113   & 1115.32 & * \\
    C-n76-6 & 75    & 6     & 3     & 75    & 1066.88 & \textbf{1060.52} & \textbf{1060.52} & 900   & 474   & \underline{1060.52} &  \\
    C-n101-4 & 100   & 4     & 4     & 100   & 1305.94 & \textbf{1302.16} & \textbf{1302.16} & 900   & 236   & \underline{1302.16} &  \\
    C-n101-5 & 100   & 5     & 4     & 100   & 1307.24 & 1306.27 & \textbf{1305.82} & 900   & 141   & \underline{1305.82} &  \\
    C-n101-6 & 100   & 6     & 4     & 100   & 1292.10 & \textbf{1284.48} & \textbf{1284.48} & 900   & 446   & 1284.48 &  \\
		\addlinespace[2pt]
    Avg.  &       &       &       &       & 912.51 & 911.11 & 910.98 & 620   & 177   & 910.92 &  \\
    \bottomrule
    \end{tabular}%
  \label{tab:set6a}%
	
\end{threeparttable}%
\end{center}
\end{adjustwidth}
\begin{adjustwidth}{-2cm}{-2cm}
\begin{center}
\begin{threeparttable}[htbp]
	\scriptsize
  \tabcolsep=3pt
  \caption{Results for Set 6b Instances ($h_s \neq 0$)}
\begin{tabular}{lccccrrrrrr@{\hspace*{0cm}}l}
    \toprule
          &       &       &       &       & \multicolumn{5}{c}{LNS-2E}               &       &  \\
		\cmidrule(lr){6-10}
    \multicolumn{1}{c}{Instance} & \multicolumn{1}{c}{C} & \multicolumn{1}{c}{S} & \multicolumn{1}{c}{T} & \multicolumn{1}{c}{CF} & \multicolumn{1}{c}{Avg. 5} & \multicolumn{1}{c}{Best 5} & \multicolumn{1}{c}{Best} & \multicolumn{1}{c}{t(s)} & \multicolumn{1}{c}{t*(s)} & \multicolumn{1}{c}{\quad BKS} &  \\
		\midrule
    \multicolumn{12}{l}{\textbf{Set 6b}}\\		
    A-n51-4 & 50    & 4     & 2     & 50    & \textbf{744.24} & \textbf{744.24} & \textbf{744.24} & 60    & 17    & 744.24 & * \\
    A-n51-5 & 50    & 5     & 2     & 50    & \textbf{811.51} & \textbf{811.51} & \textbf{811.51} & 60    & 49    & 811.52 & * \\
    A-n51-6 & 50    & 6     & 2     & 50    & \textbf{930.11} & \textbf{930.11} & \textbf{930.11} & 60    & 31    & 930.11 & * \\
    A-n76-4 & 75    & 4     & 3     & 75    & \textbf{1385.51} & \textbf{1385.51} & \textbf{1385.51} & 900   & 26    & 1385.51 & * \\
    A-n76-5 & 75    & 5     & 3     & 75    & \textbf{1519.86} & \textbf{1519.86} & \textbf{1519.86} & 900   & 71    & 1519.86 & * \\
    A-n76-6 & 75    & 6     & 3     & 75    & 1666.28 & \textbf{1666.06} & \textbf{1666.06} & 900   & 533   & 1666.06 & * \\
    A-n101-4 & 100   & 4     & 4     & 100   & 1884.48 & 1883.79 & 1883.79 & 900   & 283   & 1881.44 & * \\
    A-n101-5 & 100   & 5     & 4     & 100   & 1723.06 & 1714.58 & 1711.95 & 900   & 318   & 1709.06 &  \\
    A-n101-6 & 100   & 6     & 4     & 100   & 1795.36 & 1793.76 & 1791.44 & 900   & 100   & 1777.69 &  \\
    B-n51-4 & 50    & 4     & 2     & 50    & \textbf{653.09} & \textbf{653.09} & \textbf{653.09} & 60    & 21    & 653.09 & * \\
    B-n51-5 & 50    & 5     & 2     & 50    & \textbf{672.10} & \textbf{672.10} & \textbf{672.10} & 60    & 51    & 672.10 & * \\
    B-n51-6 & 50    & 6     & 2     & 50    & \textbf{767.13} & \textbf{767.13} & \textbf{767.13} & 60    & 58   & 767.13 & * \\
    B-n76-4 & 75    & 4     & 3     & 75    & \textbf{1094.52} & \textbf{1094.52} & \textbf{1094.52} & 900   & 44    & 1094.52 & * \\
    B-n76-5 & 75    & 5     & 3     & 75    & \textbf{1218.12} & \textbf{1218.12} & \textbf{1218.12} & 900   & 19    & 1218.13 & * \\
    B-n76-6 & 75    & 6     & 3     & 75    & 1328.90 & 1328.90 & \textbf{1326.76} & 900   & 329   & 1326.76 & * \\
    B-n101-4 & 100   & 4     & 4     & 100   & 1500.80 & \textbf{1500.55} & \textbf{1500.55} & 900   & 82    & \underline{1500.55} &  \\
    B-n101-5 & 100   & 5     & 4     & 100   & 1398.05 & 1398.05 & \textbf{1395.32} & 900   & 317   & \underline{1395.32} &  \\
    B-n101-6 & 100   & 6     & 4     & 100   & 1455.05 & 1453.54 & 1453.54 & 900   & 460   & 1450.39 &  \\
    C-n51-4 & 50    & 4     & 2     & 50    & \textbf{866.58} & \textbf{866.58} & \textbf{866.58} & 60    & 22    & 866.58 & * \\
    C-n51-5 & 50    & 5     & 2     & 50    & \textbf{943.12} & \textbf{943.12} & \textbf{943.12} & 60    & 12    & 943.12 & * \\
    C-n51-6 & 50    & 6     & 2     & 50    & \textbf{1050.42} & \textbf{1050.42} & \textbf{1050.42} & 60    & 18    & 1050.42 & * \\
    C-n76-4 & 75    & 4     & 3     & 75    & 1439.39 & \textbf{1438.96} & \textbf{1438.96} & 900   & 84    & 1438.96 & * \\
    C-n76-5 & 75    & 5     & 3     & 75    & 1745.49 & \textbf{1745.39} & \textbf{1745.39} & 900   & 281   & 1745.39 & * \\
    C-n76-6 & 75    & 6     & 3     & 75    & 1759.40 & \textbf{1756.54} & \textbf{1756.54} & 900   & 358   & 1756.54 & * \\
    C-n101-4 & 100   & 4     & 4     & 100   & 2076.36 & 2073.84 & \textbf{2064.86} & 900   & 378   & \underline{2064.86} &  \\
    C-n101-5 & 100   & 5     & 4     & 100   & 1974.39 & 1967.51 & \textbf{1964.63} & 900   & 285   & \underline{1964.63} &  \\
    C-n101-6 & 100   & 6     & 4     & 100   & 1867.45 & \textbf{1861.50} & \textbf{1861.50} & 900   & 401   & \underline{1861.50} &  \\
		\addlinespace[2pt]
    Avg.  &       &       &       &       & 1343.36 & 1342.20 & 1341.39 & 620   & 172   & 1340.57 &  \\
    \bottomrule
    \end{tabular}%
  \label{tab:set6b}%
	\quad 
\end{threeparttable}%
\end{center}
\end{adjustwidth}

To test the robustness of our algorithm we applied it also to the \ac{2E-LRPSD} instances without any further changing of operators or tuning. The results can be seen in Tables \ref{tab:lrp-n} and \ref{tab:lrp-p}. BKS are derived from \cite{Schwengerer2012, Nguyen2012a, Nguyen2012b, Contardo2012}. Also on this problem class LNS-2E is competitive, with solutions being on average within 0.6\% of the solutions found by the state-of-the-art \ac{VNS} by \cite{Schwengerer2012} (SPR).

\begin{adjustwidth}{-2cm}{-2cm}
\begin{center}
\begin{threeparttable}[htbp]
	\scriptsize
  \tabcolsep=3pt
  \centering
  \caption{Results for \acs{2E-LRPSD} Instances Set Nguyen}
    \begin{tabular}{lccrrrrrrrrrr}
    \toprule
    \multicolumn{1}{c}{} & \multicolumn{1}{c}{} & \multicolumn{1}{c}{} & \multicolumn{4}{c}{SPR}       & \multicolumn{5}{c}{LNS-2E}   \\
		\cmidrule(lr){4-7} \cmidrule(lr){8-12}
    \multicolumn{1}{c}{Instance} & \multicolumn{1}{c}{C} & \multicolumn{1}{c}{S} & \multicolumn{1}{c}{Avg. 20} & \multicolumn{1}{c}{Best 20} & \multicolumn{1}{c}{t(s)} & \multicolumn{1}{c}{t*(s)} & \multicolumn{1}{c}{Avg. 20} & \multicolumn{1}{c}{Best 20} & \multicolumn{1}{c}{Best} & \multicolumn{1}{c}{t(s)} & \multicolumn{1}{c}{t*(s)} & \multicolumn{1}{c}{BKS} \\
		\midrule
		\multicolumn{13}{l}{\textbf{Set Nguyen}}  \\
    25-5N & 25    & 5     & \textbf{80370.00} & \textbf{80370} & 76    & 3     & \textbf{80370.00} & \textbf{80370} & \textbf{80370} & 60    & 5     & 80370 \\
    25-5Nb & 25    & 5     & \textbf{64562.00} & \textbf{64562} & 91    & 0     & \textbf{64562.00} & \textbf{64562} & \textbf{64562} & 60    & 16    & 64562 \\
    25-5MN & 25    & 5     & \textbf{78947.00} & \textbf{78947} & 61    & 1     & \textbf{78947.00} & \textbf{78947} & \textbf{78947} & 60    & 6     & 78947 \\
    25-5MNb & 25    & 5     & \textbf{64438.00} & \textbf{64438} & 89    & 0     & \textbf{64438.00} & \textbf{64438} & \textbf{64438} & 60    & 4     & 64438 \\
    50-5N & 50    & 5     & \textbf{137815.00} & \textbf{137815} & 116   & 35    & \textbf{137815.00} & \textbf{137815} & \textbf{137815} & 60    & 25    & 137815 \\
    50-5Nb & 50    & 5     & 110204.40 & \textbf{110094} & 132   & 51    & 110981.85 & \textbf{110094} & \textbf{110094} & 60    & 22    & 110094 \\
    50-5MN & 50    & 5     & \textbf{123484.00} & \textbf{123484} & 125   & 41    & \textbf{123484.00} & \textbf{123484} & \textbf{123484} & 60    & 4     & 123484 \\
    50-5MNb & 50    & 5     & 105687.00 & \textbf{105401} & 202   & 48    & 105783.45 & \textbf{105401} & \textbf{105401} & 60    & 16    & 105401 \\
    50-10N & 50    & 10    & \textbf{115725.00} & \textbf{115725} & 143   & 24    & 117325.55 & \textbf{115725} & \textbf{115725} & 60    & 20    & 115725 \\
    50-10Nb & 50    & 10    & 87345.00 & \textbf{87315} & 176   & 66    & 88212.00 & 87520 & \textbf{87315} & 60    & 23    & 87315 \\
    50-10MN & 50    & 10    & \textbf{135519.00} & \textbf{135519} & 144   & 10    & 138241.35 & \textbf{135519} & \textbf{135519} & 60    & 14    & 135519 \\
    50-10MNb & 50    & 10    & \textbf{110613.00} & \textbf{110613} & 218   & 13    & 111520.80 & \textbf{110613} & \textbf{110613} & 60    & 22    & 110613 \\
    100-5N & 100   & 5     & 200685.05 & \textbf{193228} & 168   & 101   & 193806.85 & 193229 & 193229 & 900   & 239   & 193228 \\
    100-5Nb & 100   & 5     & 164508.10 & \textbf{158927} & 258   & 144   & 159064.10 & \textbf{158927} & \textbf{158927} & 900   & 315   & 158927 \\
    100-5MN & 100   & 5     & 206567.40 & \textbf{204682} & 184   & 159   & 204876.10 & \textbf{204682} & \textbf{204682} & 900   & 105   & 204682 \\
    100-5MNb & 100   & 5     & 166357.35 & \textbf{165744} & 315   & 247   & 165795.35 & \textbf{165744} & \textbf{165744} & 900   & 252   & 165744 \\
    100-10N & 100   & 10    & 214585.60 & \textbf{209952} & 223   & 167   & 216265.50 & 210799 & \textbf{209952} & 900   & 344   & 209952 \\
    100-10Nb & 100   & 10    & 155790.60 & \textbf{155489} & 352   & 251   & 161273.30 & \textbf{155489} & \textbf{155489} & 900   & 442   & 155489 \\
    100-10MN & 100   & 10    & 203798.05 & \textbf{201275} & 229   & 163   & 204396.15 & \textbf{201275} & \textbf{201275} & 900   & 324   & 201275 \\
    100-10MNb & 100   & 10    & 170791.25 & \textbf{170625} & 347   & 283   & 172202.45 & \textbf{170625} & \textbf{170625} & 900   & 268   & 170625 \\
    200-10N & 200   & 10    & 349584.15 & \textbf{345267} & 641   & 525   & 359948.65 & 350680 & 350680 & 900   & 758   & 345267 \\
    200-10Nb & 200   & 10    & 264228.90 & \textbf{256171} & 907   & 791   & 260698.20 & 257191 & 257191 & 900   & 748   & 256171 \\
    200-10MN & 200   & 10    & 332207.50 & \textbf{323801} & 453   & 441   & 329486.45 & 324279 & 324279 & 900   & 777   & 323801 \\
    200-10MNb & 200   & 10    & 292036.65 & \textbf{287076} & 944   & 843   & 297857.50 & 293339 & 290702 & 900   & 778   & 287076 \\
    \addlinespace[2pt]						
    Avg.  &       &       & 163993.75 & 161938 & 275   & 184   & 164472.98 & 162531 & 162377 & 480   & 230   & 161938 \\
    \bottomrule
    \end{tabular}%
  \label{tab:lrp-n}%
	\quad
\end{threeparttable}%
\end{center}
\end{adjustwidth}

\begin{adjustwidth}{-2cm}{-2cm}
\begin{center}
\begin{threeparttable}[htbp]
	\scriptsize
  \tabcolsep=3pt
  \centering
  \caption{Results for \acs{2E-LRPSD} Instances Set Prodhon}
    \begin{tabular}{lccrrrrrrrrrr}
    \toprule
          &       &       & \multicolumn{4}{c}{SPR}       & \multicolumn{5}{c}{LNS-2E}            &  \\
		\cmidrule(lr){4-7} \cmidrule(lr){8-12}
    \multicolumn{1}{c}{Instance} & \multicolumn{1}{c}{C} & \multicolumn{1}{c}{S} & \multicolumn{1}{c}{Avg. 20} & \multicolumn{1}{c}{Best 20} & \multicolumn{1}{c}{t(s)} & \multicolumn{1}{c}{t*(s)} & \multicolumn{1}{c}{Avg. 20} & \multicolumn{1}{c}{Best 20} & \multicolumn{1}{c}{Best} & \multicolumn{1}{c}{t(s)} & \multicolumn{1}{c}{t*(s)} & \multicolumn{1}{c}{BKS} \\
		\midrule
    \multicolumn{13}{l}{\textbf{Set Prodhon}} \\		
    20-5-1 & 20    & 5     & \textbf{89075.00} & \textbf{89075} & 63    & 2     & \textbf{89075.00} & \textbf{89075} & \textbf{89075} & 60    & 4     & 89075 \\
    20-5-1b & 20    & 5     & \textbf{61863.00} & \textbf{61863} & 83    & 0     & \textbf{61863.00} & \textbf{61863} & \textbf{61863} & 60    & 2     & 61863 \\
    20-5-2 & 20    & 5     & 84489.50 & \textbf{84478} & 62    & 11    & \textbf{84478.00} & \textbf{84478} & \textbf{84478} & 60    & 19    & 84478 \\
    20-5-2b  & 20    & 5     & 61033.80 & \textbf{60838} & 125   & 0     & \textbf{60838.00} & \textbf{60838} & \textbf{60838} & 60    & 5     & 60838 \\
    50-5-1 & 50    & 5     & 130859.30 & \textbf{130843} & 80    & 16    & 131454.00 & 131085 & \textbf{130843} & 60    & 24    & 130843 \\
    50-5-1b & 50    & 5     & 101548.00 & \textbf{101530} & 128   & 35    & 101669.20 & \textbf{101530} & \textbf{101530} & 60    & 14    & 101530 \\
    50-5-2 & 50    & 5     & \textbf{131825.00} & \textbf{131825} & 97    & 11    & 131827.00 & \textbf{131825} & \textbf{131825} & 60    & 16    & 131825 \\
    50-5-2b & 50    & 5     & \textbf{110332.00} & \textbf{110332} & 198   & 12    & \textbf{110332.00} & \textbf{110332} & \textbf{110332} & 60    & 14    & 110332 \\
    50-5-2BIS & 50    & 5     & \textbf{122599.00} & \textbf{122599} & 112   & 91    & \textbf{122599.00} & \textbf{122599} & \textbf{122599} & 60    & 3     & 122599 \\
    50-5-2bBIS & 50    & 5     & 105935.50 & \textbf{105696} & 198   & 155   & 105707.85 & \textbf{105696} & \textbf{105696} & 60    & 18    & 105696 \\
    50-5-3 & 50    & 5     & 128436.00 & \textbf{128379} & 80    & 9     & 128614.50 & \textbf{128379} & \textbf{128379} & 60    & 10    & 128379 \\
    50-5-3b & 50    & 5     & \textbf{104006.00} & \textbf{104006} & 131   & 6     & \textbf{104006.00} & \textbf{104006} & \textbf{104006} & 60    & 7     & 104006 \\
    100-5-1 & 100   & 5     & 318667.00 & 318225 & 226   & 153   & 319268.60 & 318399 & 318399 & 900   & 303   & 318134 \\
    100-5-1b & 100   & 5     & 257436.35 & 256991 & 301   & 220   & 257686.40 & 256991 & 256888 & 900   & 268   & 256878 \\
    100-5-2 & 100   & 5     & 231340.00 & \textbf{231305} & 204   & 131   & 231488.85 & \textbf{231305} & \textbf{231305} & 900   & 229   & 231305 \\
    100-5-2b & 100   & 5     & 194812.70 & 194763 & 240   & 202   & 194800.35 & 194763 & 194729 & 900   & 176   & 194728 \\
    100-5-3 & 100   & 5     & 245334.90 & 244470 & 174   & 124   & 245178.75 & \textbf{244071} & \textbf{244071} & 900   & 377   & 244071 \\
    100-5-3b & 100   & 5     & 195586.20 & 195381 & 180   & 111   & 195123.20 & \textbf{194110} & \textbf{194110} & 900   & 327   & 194110 \\
    100-10-1 & 100   & 10    & 357381.40 & 352694 & 233   & 167   & 362648.70 & 354525 & 352122 & 900   & 401   & 351243 \\
    100-10-1b & 100   & 10    & 300239.15 & 298186 & 299   & 194   & 312451.60 & 299758 & 298298 & 900   & 540   & 297167 \\
    100-10-2 & 100   & 10    & 304931.20 & 304507 & 248   & 194   & 307937.60 & 304909 & \textbf{304438} & 900   & 520   & 304438 \\
    100-10-2b & 100   & 10    & 264592.00 & 264092 & 307   & 208   & 265814.85 & 264173 & 263876 & 900   & 310   & 263873 \\
    100-10-3 & 100   & 10    & 312701.25 & 311447 & 227   & 141   & 318952.10 & 311699 & 310930 & 900   & 512   & 310200 \\
    100-10-3b & 100   & 10    & 261577.90 & 260516 & 303   & 218   & 265442.40 & 262932 & 261566 & 900   & 437   & 260328 \\
    200-10-1 & 200   & 10    & 552488.90 & 548730 & 1009  & 748   & 564159.80 & 550672 & 550672 & 900   & 725   & 548703 \\
    200-10-1b & 200   & 10    & 448095.45 & 445791 & 635   & 576   & 456952.40 & 448188 & 447113 & 900   & 692   & 445301 \\
    200-10-2 & 200   & 10    & 513673.40 & \textbf{497451} & 1158  & 832   & 499499.45 & 498486 & 498397 & 900   & 656   & 497451 \\
    200-10-2b & 200   & 10    & 432687.00 & \textbf{422668} & 730   & 696   & 428912.35 & 422967 & 422877 & 900   & 601   & 422668 \\
    200-10-3 & 200   & 10    & 529578.00 & \textbf{527162} & 970   & 903   & 568539.15 & 534271 & 533174 & 900   & 668   & 527162 \\
    200-10-3b & 200   & 10    & 404431.25 & 402117 & 592   & 557   & 425078.20 & 417686 & 417429 & 900   & 700   & 401672 \\
		\addlinespace[2pt]
    Avg.  &       &       & 245251.87 & 243599 & 313   & 224   & 248413.28 & 244720 & 244395 & 564   & 286   & 243363 \\
    \bottomrule
    \end{tabular}%
  \label{tab:lrp-p}%
	\quad
\end{threeparttable}%
\end{center}
\end{adjustwidth}

\afterpage{
\begin{figure}[htbp]
	\subfloat[number of Customers]{
	\label{box_cust}
		\includegraphics[width=0.45\textwidth]{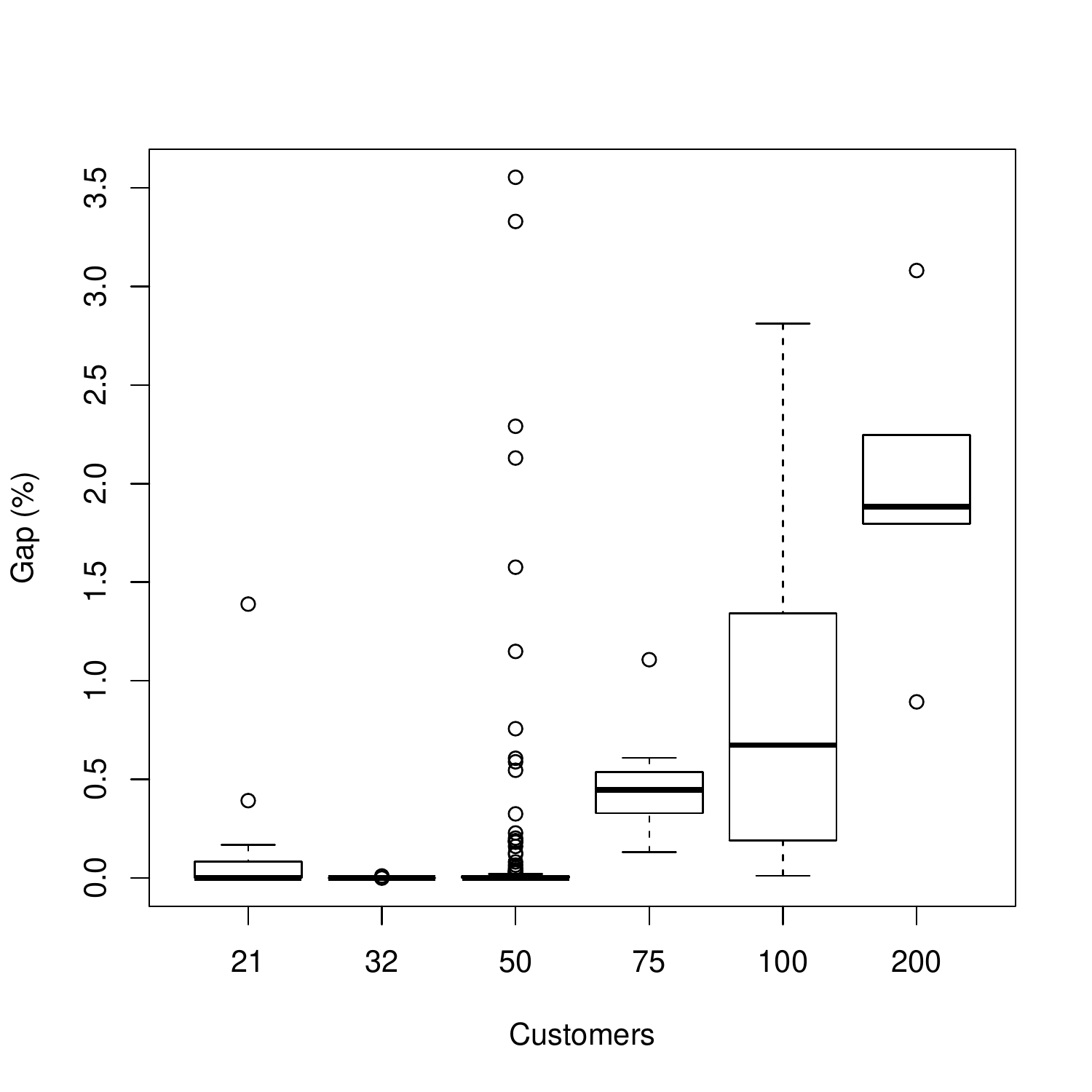}
	} 
	\hfill
	\subfloat[number of Satellites]{
	\label{box_sat}
		\includegraphics[width=0.45\textwidth]{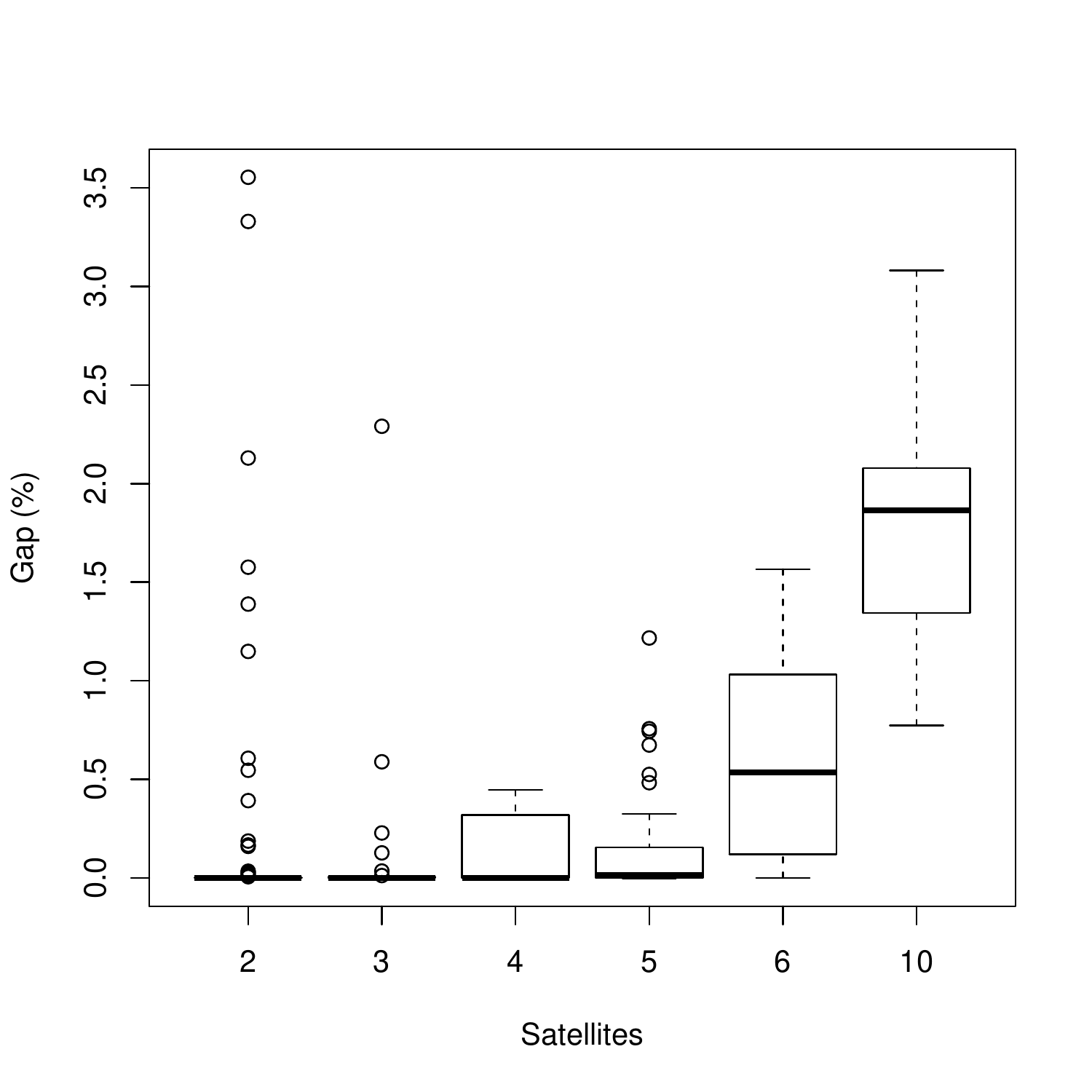}
	} 
	\hfill
	\subfloat[Set~4: Customer distribution]{
	\label{box_distr}
		\includegraphics[width=0.45\textwidth]{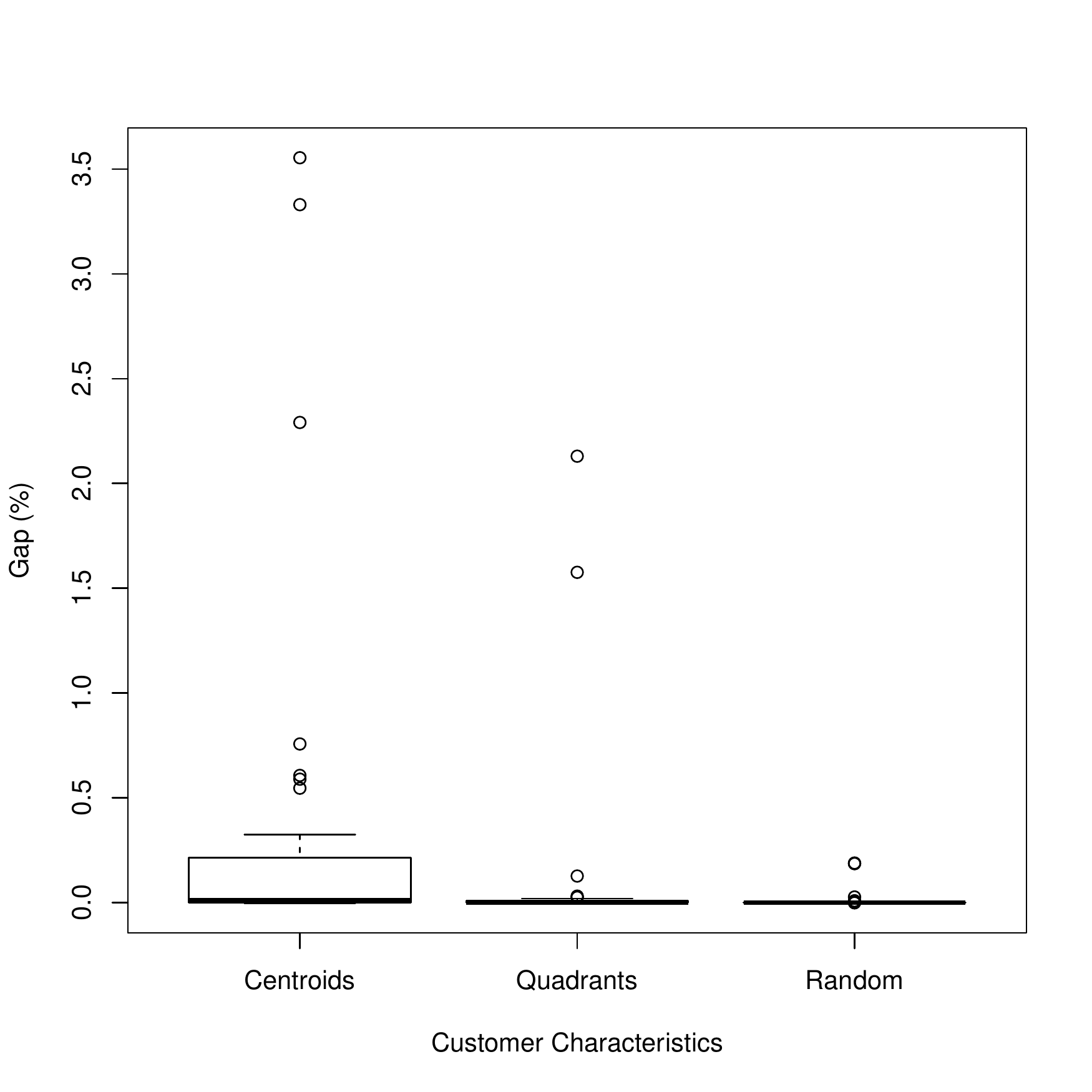}
	} 
	\hfill
	\subfloat[Set~4: Satellite distribution]{
	\label{box_distrCS}
		\includegraphics[width=0.45\textwidth]{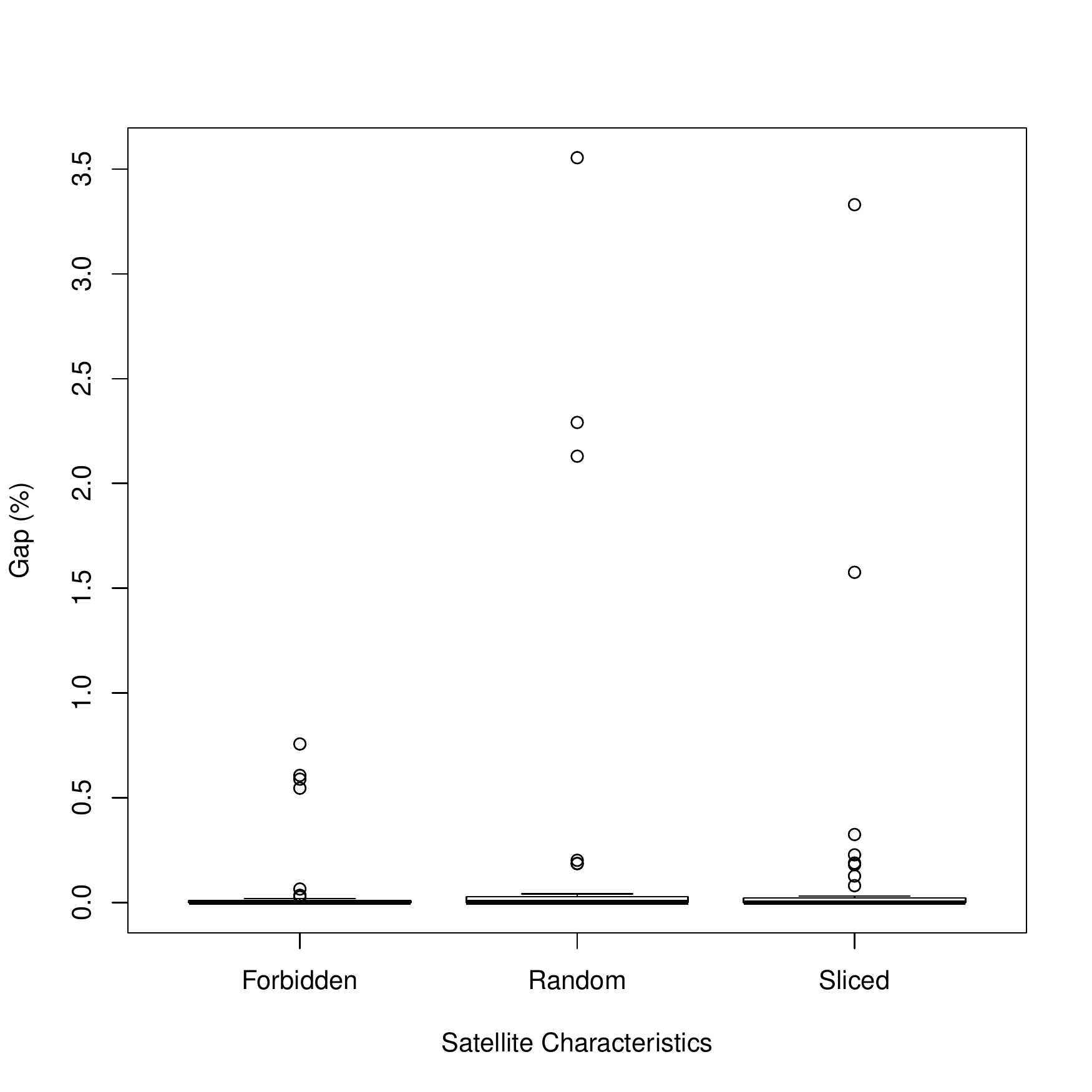}
	} 
	\caption{Boxplots\protect\footnotemark \thinspace of solution quality for instances grouped by number of customers/satellites and distribution characteristics}
	\label{boxplots}
\end{figure}

\footnotetext{The hinges depict approximately the first and third quartile of the solutions. The whiskers extend to $\pm 1.58 \frac{Q_3-Q_1}{\sqrt{n}}$. \citep{Rhelp}. See the Appendix~\ref{appendix_box} for a detailed definition.}
}

Finding high-quality solutions is increasingly difficult as the problem size grows, and different instance characteristics influence solution quality. Figure~\ref{boxplots} displays boxplots of \ac{2E-VRP} instances grouped together by number of customers (\ref{box_cust}), or number of satellites (\ref{box_sat}), similar customer distribution (\ref{box_distr}) and similar satellite distribution (\ref{box_distrCS}) using gaps of the average value of five runs to BKS. 

Problem difficulty quickly grows with the number of satellites, as well as with the number of customers, as plotted in the upper part of Figure~\ref{boxplots}. The number of samples for the different classes varies a lot: There are only 18 instances with 75 customers, but 165 instances with 50 customers, which explains the large number of outliers for those instances. 
\cite{crainic2010two} provide a detailed overview on the distribution of customers and satellites in instances of Set~4. There are three distribution patterns for customers: 
\textit{random}, with equally distributed nodes; \textit{centroids}, where more customers are located in six centroids in a central zone, and some customers closer together in four outer areas, representing suburbs. In the \textit{quadrant} pattern customers are arranged in conglomerations of higher density in each of the four quadrants. The three patterns for satellites are: \textit{random}, where satellites are randomly placed on a ring around the customers;  \textit{sliced}, with the satellites distributed more evenly on the ring around customers; and \textit{forbidden zone}: a random angle on the ring was chosen where no satellites could be used, to simulate various conditions like cities located near lakes or mountains. Figure~\ref{box_distr} shows that instances with customers located in centroids are harder to solve. On the other hand, the distribution pattern of the satellites doesn't have a large impact on solution quality.

\subsection{Graphical Example}
\label{4-38}
 Several structurally different \ac{2E-VRP} solutions can have similar objective values. We discuss and visualise this with the help of a demonstrative graphical example. Also the solution differences for an instance with or without constraining the number of city freighters per satellite are pointed out.

\begin{figure}[htbp]
	\subfloat[max. 2 CF/Sat, 1169.20*, optimal for Set 4a]{
	\label{4a-38}
		\includegraphics[width=0.3\textwidth]{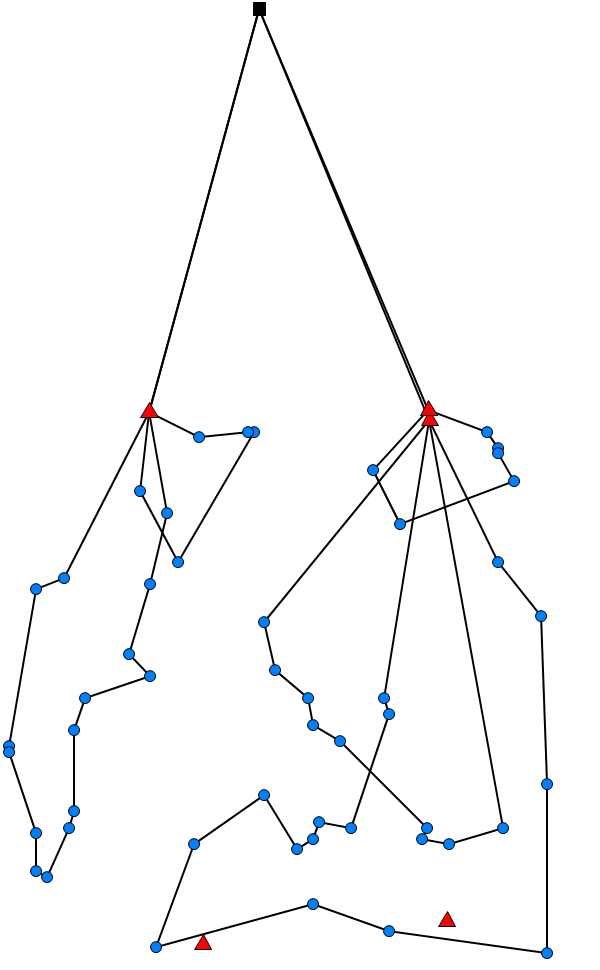}
	} 
	\hfill
	\subfloat[2 Satellites used, 1163.07*, optimal for Set 4b]{
	\label{4b-38}
		\includegraphics[width=0.3\textwidth]{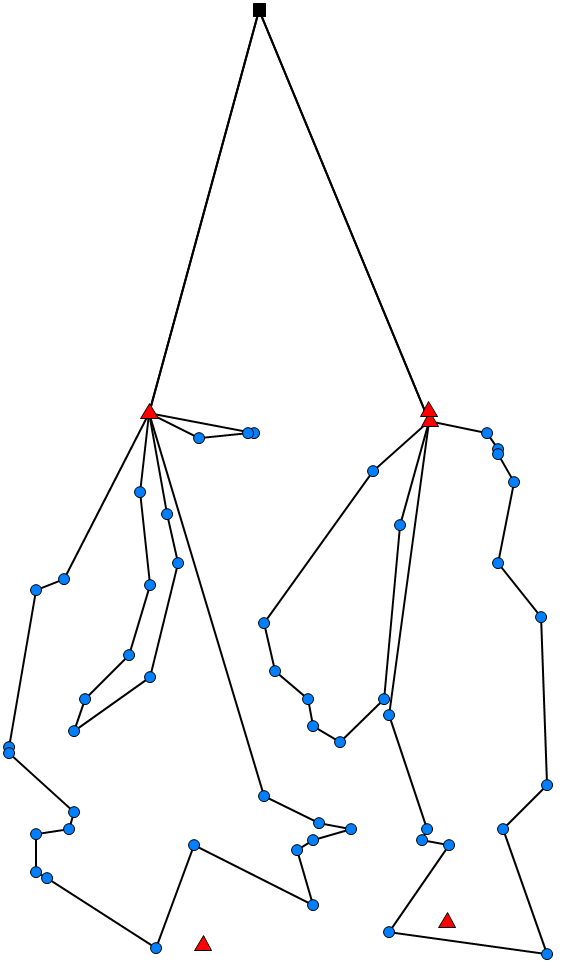}
	} 
		\hfill
	\subfloat[3 Satellites used, 1210.81]{
	\label{4b-38sub}
		\includegraphics[width=0.3\textwidth]{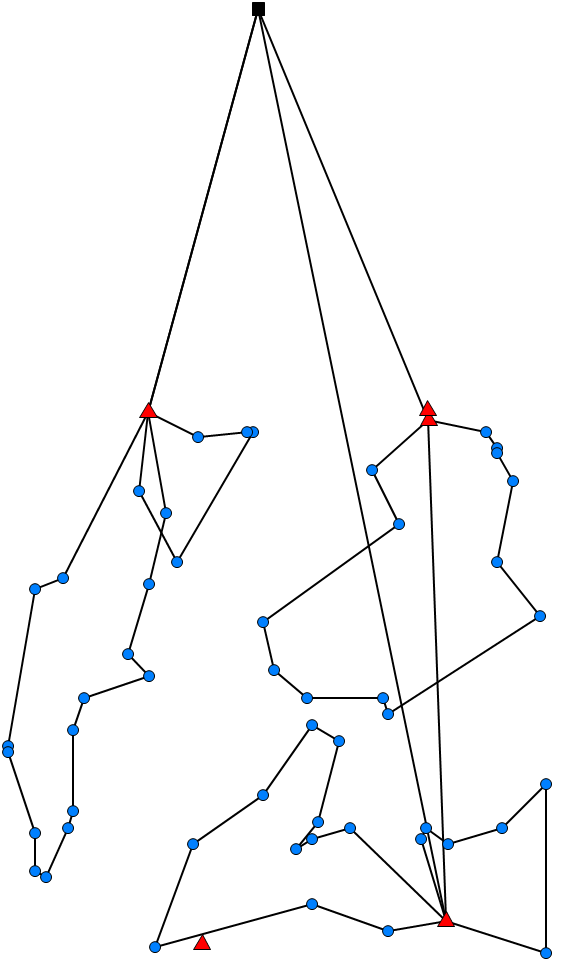}
	} 
		\hfill
	\subfloat[4 Satellites used, 1226.43]{
	\label{4b-38sub4}
		\includegraphics[width=0.3\textwidth]{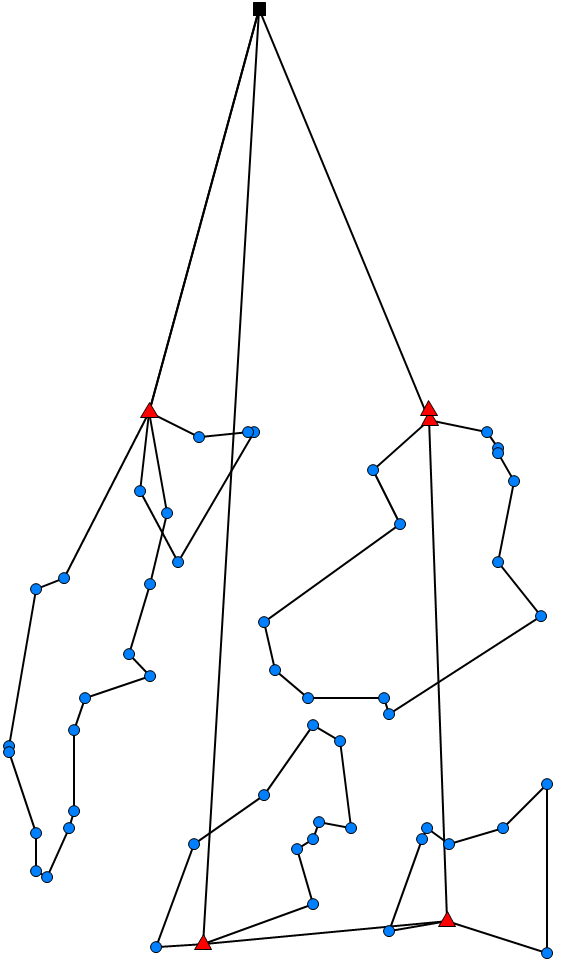}
	} 
		\hfill
	\subfloat[1 Satellite used, 1255.98]{
	\label{4b-38sub1}
		\includegraphics[width=0.3\textwidth]{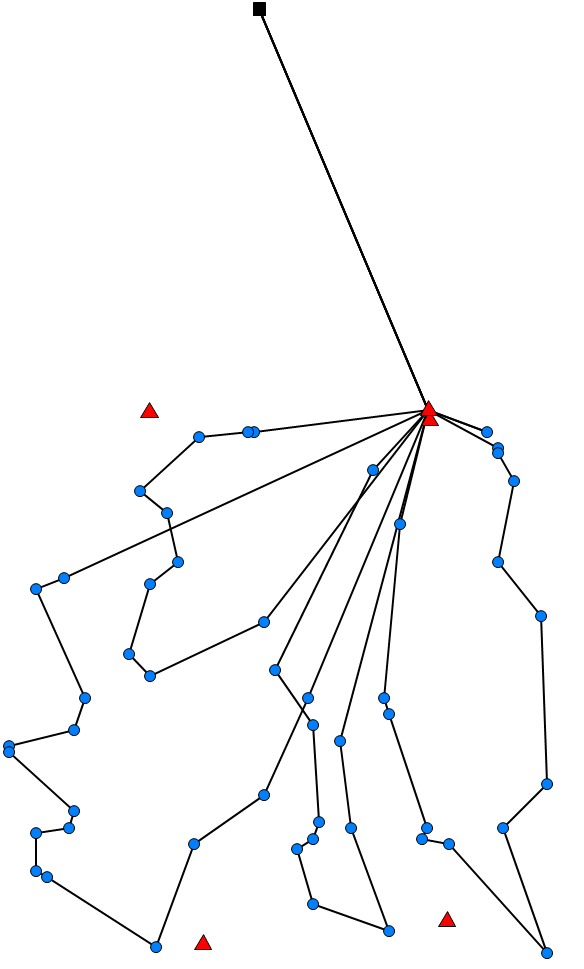}
	} 
	\caption{Different Solutions of Instance 38 from Set 4, depending on satellite openings and $v^2_s$}
	\label{Set4-38}
\end{figure}

The selection of the correct subset of satellites to use is crucial. A graphical representation of different solutions of the Set~4 instance~38 is provided in Figure~\ref{Set4-38}. The locations of the depot (square), satellites (triangles) and customers (circles) is the same for each solution. Nevertheless, the obtained vehicle routes are substantially different given different subsets of open satellites.

Figure~\ref{4a-38} represents the optimal solution to the Set~4a instance 38. A maximum of $v_s^2 = 2$ city freighters per satellite are available. The total demand of all customers sums up to 20206 units of freight. The capacity $Q^2$ of a city freighter is 5000 units. Any feasible solution needs at least $\lceil \frac{20206}{5000} \rceil = 5$ city freighters, and thus at least $\lceil 5/2 \rceil = 3$ satellites have to be used. Two city freighter routes are located on the left of the figure, and three city freighter routes on the right hand side, where two city freighters leave from the same satellite, and a third one from a close by satellite.

Considering the instance as in Set~4b, with a global number of city freighters available but no constraints on the distribution amongst satellites ($v_s^2 = v^2$), the optimal solution is displayed in Figure~\ref{4b-38}. Three city freighters can leave the same satellite, as is the case on the left hand side. Only two of the five satellites have to be used.

LNS-2E starts with the construction of routes at the second level. In early stages of the optimisation process, all customers are likely assigned to their closest satellites. Without neighbourhoods that impact the selection of satellites, the algorithm would likely be trapped in a local optimum such as the one of Figure~\ref{4b-38sub}. If partial routes originating at the bottom right satellite exist, customers may be sequentially inserted in those routes and the bottom left satellite may not be opened. The solution displayed in Figure \ref{4b-38sub4} is often obtained. It has the best cost considering only the second level, but long truck routes on the first level set this gain off, leading to a worse solution overall.

Figure~\ref{4b-38sub1} shows the best found solution if only the top right satellite is open. The cost differences from one solution to the next one are small, although the solution itself is fundamentally different. If the two satellites at the bottom are both selected for closure at the same time, then the algorithm finds the optimal solution within seconds.

We tried different strategies to evaluate the chances of a satellite to be included in the best solution: taking into account a delta evaluation on the truck route, or combining this value with the total units shipped through this satellite; or the absolute distance from the depot. We observed that closing satellites randomly is a straightforward and very simple approach, which performs quite well on average over all the different benchmark instances, whereas other techniques present advantages and disadvantages in several special cases.

For further research, we suggest to shift the cost structure towards more realistic scenarios. In the classic \ac{2E-VRP} as we considered it, the cost of large trucks and small city freighters is the same. For instance Sets~2 to 4, the capacity of a truck is 2.5 times higher than of a city freighter. For Set~5 instances this ratio gets up to more than 14, so one can safely assume that the operating cost of a truck will be higher than of a city freighter in more realistic set-ups. This has not yet been taken into account in previous publications on the \ac{2E-VRP}, and would lead to large differences between the solution costs of Figure~\ref{Set4-38}. 

\subsection{Sensitivity Analysis}
\label{sec:sensitivity}
Sensitivity analyses were conducted to evaluate the impact of major parameters and components of the method. In particular, we evaluate the impact of disabling single destroy operators or local search procedures at a time and provide the average objective value over five runs of all benchmark instances. Table \ref{tab:sensitivity} shows the average objective values for instances of each class and its average deviation (Gap (\%)) when disabling elements of the algorithm.

The elimination of the open all satellites (\emph{no open}) neighbourhood has a strong impact on solution quality. As closed satellites are only opened again in a re-start phase, the algorithm is likely to be trapped in a local optimum.  

If no satellites are forced to be closed, there is no need to open up any satellite again. In this case, solution quality deteriorates by 1.15\% on average (\emph{no close}). For some instances, satellites located very far away from the depot will not be used in the best solution, but on the second level it seems to be beneficial to use them for customers close by. The impact on the \ac{2E-LRPSD}, especially the instance Set Prodhon is quite strong, as the fixed costs of a satellite are not taken into account on the second level. If a satellite with high opening costs is located close to many customers, it will very likely be used, although it would pay off to use a cheaper one further away. A similar behaviour was discussed in Section~\ref{4-38} and Figure~\ref{Set4-38}. 

Some techniques work better on smaller instances, while others perform better on the larger instances. For some cases we even observed small improvements when an operator was not used. Not using biased node removal for example is needed to robustly find optimal solutions on the smaller instance sets (2-4) but yields improvements on the larger sets.

For local search techniques, we also observe differences on methods for small or large instance sets respectively. Removing classic 2-opt has a small impact. The repair mechanism finds already high quality single routes in terms of 2-optimality. Eliminating the inter-tour operator 2-opt* has a stronger effect on larger Set~5 and Set~6 instances, which contain more routes than the smaller instances. The relocate neighbourhood is essential for instances with more than 50 customers, whereas exchanging two nodes against one can have a negative effect on those larger instances. We still decided to keep it in the design of the algorithm, as this local search is needed to find optimal solutions for smaller instances. Of course the algorithm could be fine-tuned for specific applications or instance sizes.

\begin{center}
\begin{threeparttable}[htbp]
	\scriptsize
	\tabcolsep=3pt
  \caption{Sensitivity Analysis and contribution of individual components}
    \begin{tabular}{lrrrrrrrr}
    \toprule
    \multicolumn{9}{c}{\textbf{Sensitivity Analysis}}      \\
    Set   & base  & no related & no biased & no route & no single & no close & no open &  \\
		\midrule
    2     & 578.27 & 0.21\% & 0.02\% & 0.17\% & 0.02\% & 0.07\% & 2.07\% &  \\
    3     & 641.44 & 0.31\% & 0.06\% & 0.27\% & 0.06\% & 0.07\% & 1.66\% &  \\
    4a    & 1420.05 & 0.25\% & 0.02\% & 0.10\% & 0.04\% & 0.53\% & 1.37\% &  \\
    4b    & 1397.21 & 0.16\% & -0.01\% & 0.18\% & 0.05\% & 1.14\% & 1.39\% &  \\
    5     & 1132.02 & 0.69\% & -0.07\% & 0.31\% & 0.35\% & 0.39\% & 2.35\% &  \\
    6a    & 912.51 & 0.25\% & -0.03\% & 0.07\% & -0.01\% & 1.20\% & 1.36\% &  \\
    6b    & 1343.36 & 0.15\% & 0.04\% & 0.06\% & 0.01\% & 1.15\% & 0.97\% &  \\
		\addlinespace[2pt]
    Prodhon & 248413.28 & 0.95\% & -0.31\% & -0.55\% & -0.36\% & 4.65\% & 3.58\% &  \\
    Nguyen & 164492.82 & 0.61\% & 0.05\% & 0.28\% & 0.19\% & 1.13\% & 1.38\% &  \\
		\addlinespace[2pt]
    Avg.  & 46703.44 & 0.40\% & -0.03\% & 0.10\% & 0.04\% & 1.15\% & 1.79\% &  \\
		\midrule
		\addlinespace[8pt]
    Set   & base  & no restart & no 2opt & no 2opt* & no relocate & no swap & no swap 2 &  \\
		\midrule
    2     & 578.27 & 0.08\% & 0.04\% & 0.10\% & 0.00\% & 0.00\% & 0.00\% &  \\
    3     & 641.44 & 0.13\% & 0.05\% & 0.09\% & 0.06\% & 0.08\% & 0.18\% &  \\
    4a    & 1420.05 & 0.11\% & 0.05\% & 0.03\% & -0.01\% & 0.01\% & -0.01\% &  \\
    4b    & 1397.21 & 0.10\% & 0.03\% & 0.27\% & 0.04\% & 0.02\% & 0.03\% &  \\
    5     & 1132.02 & 0.06\% & 0.31\% & 0.53\% & 0.47\% & 0.29\% & 0.09\% &  \\
    6a    & 912.51 & 0.11\% & 0.02\% & 0.15\% & 0.24\% & -0.03\% & 0.02\% &  \\
    6b    & 1343.36 & 0.07\% & 0.05\% & 0.17\% & 0.12\% & 0.05\% & -0.01\% &  \\
		\addlinespace[2pt]
    Prodhon & 248413.28 & 0.43\% & -0.29\% & 0.30\% & 0.56\% & -0.51\% & 0.00\% &  \\
    Nguyen & 164492.82 & -0.08\% & -0.20\% & 0.21\% & 0.40\% & 0.22\% & 0.03\% &  \\
		\addlinespace[2pt]
    Avg.  & 46703.44 & 0.11\% & 0.01\% & 0.21\% & 0.21\% & 0.01\% & 0.04\% &  \\
    \bottomrule
    \end{tabular}%
  \label{tab:sensitivity}%
\end{threeparttable}%
\end{center}

\section{Conclusion}
\label{conclusion}
We presented a very simple and fast \ac{LNS} heuristic for the \ac{2E-VRP} and \ac{2E-LRPSD}. LNS-2E makes use of one repair operator and only a few destroy operators. The proposed method finds solutions of higher quality than existing algorithms, while being fast and conceptually simpler. The impact of various parameters and design choices was highlighted. Meta-calibration techniques were used to set the parameters to good values, which were subsequently verified during sensitivity analyses. Different techniques were attempted to open or close satellites, and thus explore various combinations of design choices. At the end, a simple randomised approach for fixing satellites, assorted with a minimum number of iterations without change of this decision led to good and robust results on a wide range of benchmark instances. LNS-2E was able to improve 18 best known solutions on the 49 \ac{2E-VRP} instances for which no proven optimal solution exists so far. Having resolved the inconsistencies on the different sets of benchmark instances used in literature paves the way for future research on this topic, which will focus on solving rich city logistics problems, and shifting the cost structure to a more realistic scenario. It will be interesting to examine the implications of using higher operating costs for larger trucks than for smaller city freighters, approaching more realistic and real-life transportation problems.

\section*{Online Resources}
All necessary data can be found in the online section at \url{https://www.univie.ac.at/prolog/research/TwoEVRP}. All instances have been transformed into a uniform format and can be downloaded. Set~4 instances used to have negative and real x/y coordinates (with a maximum of two positions after decimal space). We added a fixed factor to shift them only to be positive and multiplied them by 100 to be able to use positive integers. This does not change the solution, but note that the objective should be adjusted by a factor 100.
We also provide detailed results on the new found best known solutions, both in human readable text and a graphical representation.

\section*{Acknowledgement}
This work is partially funded by the Austrian Climate and Energy Fund within the ``Electric Mobility Flagship projects'' program under grant 834868 (project VECEPT). The financial support by the Austrian Federal Ministry of Science, Research and Economy and the National Foundation for Research, Technology and Development is gratefully acknowledged.

We would like to thank Roberto Baldacci, Vera Hemmelmayr, Sandro Pirkwieser, Martin Schwengerer and Gianfranco Guastaroba for providing us detailed information, and/or solutions on the instances solved by their methods.

We would also like to thank the two anonymous reviewers for their helpful and constructive comments which helped us to improve the manuscript.

\section{Appendix}
\label{appendix}

\subsection{Nomenclature of Set~3 instances}
\begin{center}
\begin{threeparttable}[htbp]
	\small
  \caption{Set~3 instances with 50 customers: identical except for depot coordinates}
    \begin{tabular}{rcrc}
    \toprule
		\multicolumn{2}{c}{Set 3b\tnote{1}} &  \multicolumn{2}{c}{Set 3c\tnote{2}} \\
    \multicolumn{2}{c}{depot at (0,0)} &   \multicolumn{2}{c}{depot at (30,40)} \\
		\cmidrule(lr){1-2} \cmidrule(lr){3-4}
    E-n51-k5- & 12-18 &   E-n51-k5- & 13-19 \\
          & 12-41 &        & 13-42 \\
          & 12-43 &        & 13-44 \\
          & 39-41 &        & 40-42 \\
          & 40-41 &        & 41-42 \\
          & 40-43 &        & 41-44 \\
    \bottomrule
    \end{tabular}%
  \label{tab:set3depot}%
	\begin{tablenotes}\tiny
		\item[1] included in \cite{hemmelmayr2013email}
		\item[2] included in \cite{orlib}
	\end{tablenotes}
\end{threeparttable}%
\end{center}

 \cite{Hemmelmayr2012} report a best solution with total costs of 690.59 for instance E-n51-k5-s\textbf{12-18}, which corresponds exactly to the solution found by our algorithm. On the other hand \cite{Jepsen2012} report an objective value of 560.73 for an instance of that name. This corresponds exactly to the objective value found by our algorithm for the instance E-n51-k5-s\textbf{13-19} from \cite{orlib}. So we assume \cite{Jepsen2012} used the same instances as are provided from \cite{orlib}, but the IDs of satellites in the names have to be increased by one. Both versions exist in literature, as we described in Section~\ref{instances}. The only difference between Sets~3b and 3c is the location of the depot. Table~\ref{tab:set3depot} shows, which instances correspond to each other, apart from depot coordinates.

\subsection{Definition of boxplots in Figure~\ref{boxplots}}
\label{appendix_box}
The two hinges are versions of the first ($Q_1$) and third quartile ($Q_3$), i.e., close to quantile (x,~c(1,3)/4). The hinges equal the quartiles for odd n (where $n\leftarrow~length(x)$) and differ for even n. Whereas the quartiles only equal ob\-ser\-va\-tions for $n = (1\mod~4)$, the hinges do so additionally for $n=(2\mod~4)$, and are in the middle of two ob\-ser\-va\-tions otherwise. They are based on asymptotic normality of the median and roughly equal sample sizes for the two medians being compared, and are said to be rather insensitive to the underlying distributions of the samples.
The notches extend to $\pm~1.58\frac{IQR}{\sqrt{n}}$, where Interquartile Range $IQR=Q_3-Q_1$. This gives roughly a 95\% confidence interval for the difference in two medians. \citep{Rhelp}

\subsection{\ac{2E-LRPSD} instances}
\label{sec:app-LRPinst}
The instances for the \ac{2E-LRPSD} were downloaded at the homepage\footnote{\url{http://prodhonc.free.fr/Instances/instances0_us.htm}} of Caroline Prodhon. If the description file is renamed to ``.doc'' it can be opened in human readable format by Word. We found some small typos in the description: At the end of the Prodhon files first the fixed cost of a second level vehicle (F2) is given, then the fixed cost of a truck on first level (F1).

The description of the cost structure also seems to contain an error: According to the description, second level vehicles operate at higher cost per distance than first level trucks, which should be the other way round, obviously. This is also the way we treated the instances, and we believe previous authors did, too. The cost linking point A to point B was calculated as shown in Table~\ref{tab:distMatrix}. Please note that first level vehicles do not operate at exactly twice the Euclidean distance, due to the use of the ceil function \textit{after} multiplication by factor 2.

\begin{table}[htbp]
	\scriptsize
	\tabcolsep=3pt
  \centering
  \caption{Distance Matrix Calculation for the \ac{2E-LRPSD}}
    \begin{tabular}{rll}
    \toprule
          & \multicolumn{2}{c}{Instance set} \\
    \cmidrule(lr){2-3}
          & \multicolumn{1}{c}{Prodhon} & \multicolumn{1}{c}{Nguyen} \\
    first level & $\lceil \sqrt{(x_A-x_B)^2 + (y_A-y_B)^2}*100*2 \rceil$  & $\lceil \sqrt{(x_A-x_B)^2 + (y_A-y_B)^2}*10*2 \rceil$ \\
		\addlinespace[2pt]
    second level & $\lceil \sqrt{(x_A-x_B)^2 + (y_A-y_B)^2}*100 \rceil$  & $\lceil \sqrt{(x_A-x_B)^2 + (y_A-y_B)^2}*10 \rceil$ \\
    \bottomrule
    \end{tabular}%
  \label{tab:distMatrix}%
\end{table}%

In the instance file 200-10-3b, the capacity of first level vehicles is missing, and thus we used 5000 (as this is the value used for all the other Prodhon instance files). Some customers in the \ac{2E-LRPSD} instances have a demand of 0. This case was not explicitly dealt with in the papers: our algorithm still plans an itinerary for a city freighter which will visit the customer, but does not deliver any quantity of goods.

\section*{Acronyms}
\begin{acronym}
	\setlength{\itemsep}{-\parsep}																		
		\acro{2E-VRP}[2EVRP]{two-echelon vehicle routing problem}
		\acro{2E-LRP}[2ELRP]{two-echelon location routing problem}
		\acro{2E-LRPSD}[2ELRPSD]{two-echelon location routing problem with single depot}
		\acro{AIT}{Austrian Institute of Technology}
		\acro{ALNS}{adaptive large neighbourhood search}
		\acro{BKS}{best known solution}
		\acro{cf}[cf.\ ]{confer}
		\acro{CMA-ES}{covariance matrix adaptation evolution strategy}
		\acro{Cust}[Cust.\ ]{Customer}
		\acro{CVRP}{capacitated vehicle routing problem}
		\acro{distrib}[distrib.\ ]{distribution}
		\acrodef{eg}[e.g.] {example given}
		\acrodef{etc}[etc.]{et cetera}
		\acrodef{ie}[i.e.]{id est}
		\acro{E-VRPTW}{electric vehicle routing problem with time windows and recharging stations} 
		\acro{Inst}[Inst.]{Instance}
		\acro{LNS}{large neighbourhood search}
		\acro{LRP}{location routing problem}
		\acro{MD-VRP}[MDVRP]{mul\-ti-de\-pot ve\-hi\-cle rou\-ting pro\-blem}
		\acro{MIP}{mixed integer program}
		\acro{NP}{non-deterministic polynomial-time}
		\acro{p}[p.]{page}
		\acro{Sat}[Sat.]{Satellite}
		\acro{SD-VRP}[SDVRP]{split delivery vehicle routing problem}
		\acro{Sol}[Sol.]{Solution}
		\acro{T}[T.]{Time}
		\acro{TSP}{tra\-vel\-ling sales\-man prob\-lem}
		\acro{VLNS}{very large neighbourhood search}
		\acro{VNS}{variable neighbourhood search}
		\acro{vrp}[VRP]{vehicle routing problem}
		\acrodef{Vrp}[VRP]{Vehicle routing problem}
\end{acronym}	

\section*{Bibliography}
\bibliographystyle{abbrvnat}
\bibliography{library}

\begin{thebibliography}{35}
\providecommand{\natexlab}[1]{#1}
\providecommand{\url}[1]{\texttt{#1}}
\expandafter\ifx\csname urlstyle\endcsname\relax
  \providecommand{\doi}[1]{doi: #1}\else
  \providecommand{\doi}{doi: \begingroup \urlstyle{rm}\Url}\fi

\bibitem[Baldacci(2013)]{baldacci2013email}
R.~Baldacci.
\newblock personal communication.
\newblock email, May 20th 2013.

\bibitem[Baldacci et~al.(2013)Baldacci, Mingozzi, Roberti, and
  Wolfler~Calvo]{Baldacci2013}
R.~Baldacci, A.~Mingozzi, R.~Roberti, and R.~Wolfler~Calvo.
\newblock {An Exact Algorithm for the Two-Echelon Capacitated Vehicle Routing
  Problem}.
\newblock \emph{Operations Research}, 61\penalty0 (2):\penalty0 298--314, 2013.
\newblock \doi{10.1287/opre.1120.1153}.

\bibitem[Beasley(2014)]{orlib}
J.~E. Beasley.
\newblock {OR-Library}, accessed Oct. 6th 2014.
\newblock URL
  \url{http://people.brunel.ac.uk/~mastjjb/jeb/orlib/files/vrp2e.rar}.

\bibitem[Boccia et~al.(2011)Boccia, Crainic, Sforza, and
  Sterle]{boccia2011location}
M.~Boccia, T.~G. Crainic, A.~Sforza, and C.~Sterle.
\newblock {Location-Routing Models for Designing a Two-Echelon Freight
  Distribution System}.
\newblock Technical report, Cirrelt, 2011.

\bibitem[Contardo et~al.(2012)Contardo, Hemmelmayr, and Crainic]{Contardo2012}
C.~Contardo, V.~C. Hemmelmayr, and T.~G. Crainic.
\newblock {Lower and upper bounds for the two-echelon capacitated
  location-routing problem}.
\newblock \emph{Computers \& Operations Research}, 39:\penalty0 3185--3199,
  2012.

\bibitem[Cordeau et~al.(2002)Cordeau, Gendreau, Laporte, Potvin, and
  Semet]{Cordeau2002}
J.-F. Cordeau, M.~Gendreau, G.~Laporte, J.-Y. Potvin, and F.~Semet.
\newblock {A guide to vehicle routing heuristics}.
\newblock \emph{Journal of the Operational Research Society}, 53\penalty0
  (5):\penalty0 512--522, 2002.
\newblock \doi{10.1057/palgrave/jors/2601319}.

\bibitem[Crainic et~al.(2004)Crainic, Ricciardi, and
  Storchi]{crainic2004advanced}
T.~G. Crainic, N.~Ricciardi, and G.~Storchi.
\newblock {Advanced Freight Transportation Systems for Congested Urban Areas}.
\newblock \emph{Transportation Research Part C: Emerging Technologies},
  12\penalty0 (2):\penalty0 119--137, 2004.
\newblock \doi{10.1016/j.trc.2004.07.002}.

\bibitem[Crainic et~al.(2009)Crainic, Ricciardi, and
  Storchi]{crainic2009models}
T.~G. Crainic, N.~Ricciardi, and G.~Storchi.
\newblock {Models for Evaluating and Planning City Logistics Systems}.
\newblock \emph{Transportation Science}, 43\penalty0 (4):\penalty0 432--454,
  2009.
\newblock \doi{10.1287/trsc.1090.0279}.

\bibitem[Crainic et~al.(2010)Crainic, Perboli, Mancini, and
  Tadei]{crainic2010two}
T.~G. Crainic, G.~Perboli, S.~Mancini, and R.~Tadei.
\newblock {Two-Echelon Vehicle Routing Problem: a Satellite Location Analysis}.
\newblock \emph{Procedia - Social and Behavioral Sciences}, 2\penalty0
  (3):\penalty0 5944--5955, 2010.
\newblock \doi{10.1016/j.sbspro.2010.04.009}.

\bibitem[Crainic et~al.(2011)Crainic, Mancini, Perboli, and
  Tadei]{Crainic2011b}
T.~G. Crainic, S.~Mancini, G.~Perboli, and R.~Tadei.
\newblock {Multi-Start Heuristics for the Two-Echelon Vehicle Routing Problem}.
\newblock In P.~Merz and J.-K. Hao, editors, \emph{Evolutionary Computation in
  Combinatorial Optimization}, volume 6622, pages 179--190. Springer Berlin
  Heidelberg, Berlin Heidelberg, 2011.

\bibitem[Croes(1958)]{croes1958method}
G.~A. Croes.
\newblock {A method for solving traveling-salesman problems}.
\newblock \emph{Operations Research}, 6\penalty0 (6):\penalty0 791--812, 1958.
\newblock \doi{10.1287/opre.6.6.791}.

\bibitem[Cuda et~al.(2015)Cuda, Guastaroba, and Speranza]{Cuda2014a}
R.~Cuda, G.~Guastaroba, and M.~G. Speranza.
\newblock {A Survey on Two-Echelon Routing Problems}.
\newblock \emph{Computers \& Operations Research}, 55:\penalty0 185--199, 2015.
\newblock \doi{10.1016/j.cor.2014.06.008}.

\bibitem[Hansen(2006)]{hansen2006}
N.~Hansen.
\newblock The {CMA} evolution strategy: a comparing review.
\newblock In J.~Lozano, P.~Larranaga, I.~Inza, and E.~Bengoetxea, editors,
  \emph{Towards a new evolutionary computation. Advances on estimation of
  distribution algorithms}, pages 75--102. Springer Berlin Heidelberg, 2006.

\bibitem[Hemmelmayr(2013)]{hemmelmayr2013email}
V.~C. Hemmelmayr.
\newblock personal communication.
\newblock email, March 21st 2013.

\bibitem[Hemmelmayr et~al.(2012)Hemmelmayr, Cordeau, and
  Crainic]{Hemmelmayr2012}
V.~C. Hemmelmayr, J.-F. Cordeau, and T.~G. Crainic.
\newblock {An Adaptive Large Neighborhood Search Heuristic for Two-Echelon
  Vehicle Routing Problems Arising in City Logistics}.
\newblock \emph{Computers \& Operations Research}, 39\penalty0 (12):\penalty0
  3215--3228, 2012.
\newblock \doi{10.1016/j.cor.2012.04.007}.

\bibitem[Jacobsen and Madsen(1980)]{Jacobsen1980}
S.~Jacobsen and O.~Madsen.
\newblock {A Comparative Study of Heuristics for a Two-Level Routing-Location
  Problem}.
\newblock \emph{European Journal of Operational Research}, 5\penalty0
  (6):\penalty0 378--387, 1980.
\newblock \doi{10.1016/0377-2217(80)90124-1}.

\bibitem[Jepsen et~al.(2012)Jepsen, Spoorendonk, and Ropke]{Jepsen2012}
M.~Jepsen, S.~Spoorendonk, and S.~Ropke.
\newblock {A Branch-and-Cut Algorithm for the Symmetric Two-Echelon Capacitated
  Vehicle Routing Problem}.
\newblock \emph{Transportation Science}, 47\penalty0 (1):\penalty0 23--37,
  2012.
\newblock \doi{10.1287/trsc.1110.0399}.

\bibitem[Laporte(1988)]{Laporte1988}
G.~Laporte.
\newblock {Location routing problems}.
\newblock In B.~L. Golden and A.~A. Assad, editors, \emph{Vehicle Routing:
  Methods and Studies}, pages 163--197. Elsevier Science Publishers B.V.,
  North-Holland, 1988.

\bibitem[Laporte and Nobert(1988)]{LaporteNobert1988}
G.~Laporte and Y.~Nobert.
\newblock A vehicle flow model for the optimal design of a two-echelon
  distribution system.
\newblock In H.~A. Eiselt and G.~Pederzoli, editors, \emph{Advances in
  Optimization and Control: Proceedings of the Conference ``Optimization Days
  86'' Held at Montreal, Canada, April 30 -- May 2, 1986}, pages 158--173,
  Berlin, Heidelberg, 1988. Springer Berlin Heidelberg.
\newblock \doi{10.1007/978-3-642-46629-8_11}.

\bibitem[Madsen(1983)]{Madsen1983}
O.~B. Madsen.
\newblock {Methods for solving combined two level location-routing problems of
  realistic dimensions}.
\newblock \emph{European Journal of Operational Research}, 12\penalty0
  (3):\penalty0 295--301, mar 1983.
\newblock \doi{10.1016/0377-2217(83)90199-6}.

\bibitem[Nguyen et~al.(2012{\natexlab{a}})Nguyen, Prins, and
  Prodhon]{Nguyen2012a}
V.~P. Nguyen, C.~Prins, and C.~Prodhon.
\newblock {Solving the two-echelon location routing problem by a GRASP
  reinforced by a learning process and path relinking}.
\newblock \emph{European Journal of Operational Research}, 216\penalty0
  (1):\penalty0 113--126, 2012{\natexlab{a}}.
\newblock \doi{10.1016/j.ejor.2011.07.030}.

\bibitem[Nguyen et~al.(2012{\natexlab{b}})Nguyen, Prins, and
  Prodhon]{Nguyen2012b}
V.~P. Nguyen, C.~Prins, and C.~Prodhon.
\newblock {A multi-start iterated local search with tabu list and path
  relinking for the two-echelon location-routing problem}.
\newblock \emph{Engineering Applications of Artificial Intelligence},
  25\penalty0 (1):\penalty0 56--71, 2012{\natexlab{b}}.
\newblock \doi{10.1016/j.engappai.2011.09.012}.

\bibitem[Perboli and Tadei(2010)]{perboli2010new}
G.~Perboli and R.~Tadei.
\newblock {New Families of Valid Inequalities for the Two-Echelon Vehicle
  Routing Problem}.
\newblock \emph{Electronic Notes in Discrete Mathematics}, 36:\penalty0
  639--646, 2010.

\bibitem[Perboli et~al.(2011)Perboli, Tadei, and Vigo]{Perboli2011}
G.~Perboli, R.~Tadei, and D.~Vigo.
\newblock {The Two-Echelon Capacitated Vehicle Routing Problem: Models and
  Math-Based Heuristics}.
\newblock \emph{Transportation Science}, 45\penalty0 (3):\penalty0 364--380,
  2011.
\newblock \doi{10.1287/trsc.1110.0368}.

\bibitem[Pirkwieser and Raidl(2010)]{Pirkwieser2010}
S.~Pirkwieser and G.~R. Raidl.
\newblock {Variable Neighborhood Search coupled with ILP-based Very Large
  Neighborhood Searches for the (Periodic) Location-Routing Problem}.
\newblock In \emph{Lecture Notes in Computer Science (including subseries
  Lecture Notes in Artificial Intelligence and Lecture Notes in
  Bioinformatics)}, volume 6373 LNCS, pages 174--189, Berlin, Heidelberg, 2010.
  Springer Berlin Heidelberg.
\newblock \doi{10.1007/978-3-642-16054-7_13}.

\bibitem[Pisinger and Ropke(2010)]{pisinger2010large}
D.~Pisinger and S.~Ropke.
\newblock {Large Neighborhood Search}.
\newblock In M.~Gendreau and J.-Y. Potvin, editors, \emph{Handbook of
  Metaheuristics}, volume 146, pages 399--419. Springer, 2010.

\bibitem[Santos et~al.(2013)Santos, {Salles da Cunha}, and Mateus]{Santos2013}
F.~A. Santos, A.~{Salles da Cunha}, and G.~R. Mateus.
\newblock {Branch-and-price algorithms for the Two-Echelon Capacitated Vehicle
  Routing Problem}.
\newblock \emph{Optimization Letters}, 7:\penalty0 1537--1547, 2013.
\newblock \doi{10.1007/s11590-012-0568-3}.

\bibitem[Santos et~al.(2014)Santos, Mateus, and {Salles da Cunha}]{Santos2014}
F.~A. Santos, G.~R. Mateus, and A.~{Salles da Cunha}.
\newblock {A Branch-and-Cut-and-Price Algorithm for the Two-Echelon Capacitated
  Vehicle Routing Problem}.
\newblock \emph{Transportation Science}, 2014.
\newblock \doi{10.1287/trsc.2013.0500}.

\bibitem[Schrimpf et~al.(2000)Schrimpf, Schneider, Stamm-Wilbrandt, and
  Dueck]{Schrimpf2000}
G.~Schrimpf, J.~Schneider, H.~Stamm-Wilbrandt, and G.~Dueck.
\newblock {Record Breaking Optimization Results Using the Ruin and Recreate
  Principle}.
\newblock \emph{Journal of Computational Physics}, 159:\penalty0 139--171,
  2000.
\newblock \doi{10.1006/jcph.1999.6413}.

\bibitem[Schwengerer et~al.(2012)Schwengerer, Pirkwieser, and
  Raidl]{Schwengerer2012}
M.~Schwengerer, S.~Pirkwieser, and G.~R. Raidl.
\newblock A variable neighborhood search approach for the two-echelon
  location-routing problem.
\newblock In J.-K. Hao and M.~Middendorf, editors, \emph{Evolutionary
  Computation in Combinatorial Optimization}, volume 7245 of \emph{Lecture
  Notes in Computer Science}, pages 13--24. Springer Berlin Heidelberg, 2012.
\newblock \doi{10.1007/978-3-642-29124-1_2}.

\bibitem[Shaw(1998)]{shaw1998using}
P.~Shaw.
\newblock {Using Constraint Programming and Local Search Methods to Solve
  Vehicle Routing Problems}.
\newblock In M.~Maher and J.-F. Puget, editors, \emph{Principles and Practice
  of Constraint Programming}, volume 1520 of \emph{Lecture Notes in Computer
  Science}, pages 417--431. Springer Berlin Heidelberg, 1998.
\newblock \doi{10.1007/3-540-49481-2\_30}.

\bibitem[{The R Foundation for Statistical Computing}(2014)]{Rhelp}
{The R Foundation for Statistical Computing}.
\newblock R help documentation on boxplot.stats, version 3.1.2, 2014.
\newblock URL
  \url{http://web.mit.edu/r/r_v3.0.1/lib/R/library/grDevices/html/boxplot.stats.html}.
\newblock retrieved Feb. 20th 2015.

\bibitem[Toth and Vigo(2003)]{Toth2003}
P.~Toth and D.~Vigo.
\newblock {The Granular Tabu Search and its Application to the Vehicle-Routing
  Problem}.
\newblock \emph{INFORMS Journal on Computing}, 15\penalty0 (4):\penalty0
  333--346, 2003.

\bibitem[Vidal et~al.(2013)Vidal, Crainic, Gendreau, and Prins]{Vidal2013}
T.~Vidal, T.~G. Crainic, M.~Gendreau, and C.~Prins.
\newblock {Heuristics for Multi-Attribute Vehicle Routing Problems: A Survey
  and Synthesis}.
\newblock \emph{European Journal of Operational Research}, 231\penalty0
  (1):\penalty0 1--21, 2013.

\bibitem[Zeng et~al.(2014)Zeng, Xu, Xu, and Shao]{Zeng2014a}
Z.-y. Zeng, W.-s. Xu, Z.-y. Xu, and W.-h. Shao.
\newblock {A Hybrid GRASP+VND Heuristic for the Two-Echelon Vehicle Routing
  Problem Arising in City Logistics}.
\newblock \emph{Mathematical Problems in Engineering}, 2014:\penalty0 1--11,
  2014.
\newblock \doi{10.1155/2014/517467}.

\end{thebibliography}

\end{document}